\documentclass[superscriptaddress,aps,prd,showpacs,twocolumn]{revtex4}
\pdfoutput=1

\usepackage[english]{babel}
\usepackage[utf8]{inputenc}

\usepackage[dvips]{graphicx}
\usepackage[usenames]{xcolor}
\usepackage{amsmath}
\usepackage{amscd}
\usepackage{listings}
\usepackage{enumerate}
\usepackage{amssymb}
\usepackage{amsthm}
\usepackage{syntonly}
\usepackage{fancyhdr}

\DeclareFixedFont{\garde}{OT1}{pag}{m}{n}{11pt}

\usepackage{layout}
\usepackage{physics}
\usepackage{slashed}

\usepackage{relsize}
\DeclareFixedFont{\garde}{OT1}{pag}{m}{n}{11pt}
\usepackage[mathscr]{euscript}
\usepackage{dsfont}
\usepackage{amsfonts}

\usepackage{tabstackengine}
\stackMath
\setstackgap{L}{15pt}
\setstacktabbedgap{4pt}
\def\lrgap{\kern-1pt}
\def\doublet#1{\left(\lrgap\Vectorstack{#1}\lrgap\right)}

\usepackage[pdftex]{hyperref}
\hypersetup{colorlinks,linkcolor=blue,citecolor=red}

\begin{document}

\newcommand{\Lag}{\mathcal L}
\newcommand{\ddGamma}{\mathbb{I}\hspace{-1.mm}\Gamma}
\newcommand{\Reals}{\mathds R}
\newcommand{\Complex}{\mathds C}
\newcommand{\Q}{\mathrm Q}
\newcommand{\rbvarphi}{{\fontsize{2.3225ex}{2.787ex}
\selectfont\raisebox{0.5ex}{$\boldsymbol{\varphi}$}}}
\newcommand{\nchi}{\mathscr X}
\newcommand{\ttK}{\mathtt K}
\newcommand{\ttX}{\mathtt X}
\newcommand{\ttN}{\mathtt N}
\newcommand{\ttV}{\mathtt V}
\newcommand{\ttE}{\mathtt E}

\bibliographystyle{unsrt}

\title{Cosmological Unification, Dark Energy and the Origin of Neutrino Mass}
\author{J.G. Salazar-Arias}
\email{jsalazar@fis.cinvestav.mx}
\author{A. P\'erez-Lorenzana}
\email{aplorenz@fis.cinvestav.mx}
\affiliation{Centro de Investigaci\'on y de Estudios Avanzados del I.P.N.
 Apartado Postal 14-740,07000 Mexico City, Mexico }
\date{\today}

\begin{abstract}
We suggest that quintessential vacuum energy could be the source of right 
handed neutrino masses that feed the seesaw mechanism, which may provide observed 
small masses to light standard neutrinos. This idea is naturally implemented in 
the Cosmological Unification model based on the global $SO(1,1)$ symmetry, where 
early inflation and late accelerated expansion of the Universe are driven by 
the degrees of freedom of a doublet scalar field. In this model, the $SO(1,1)$ 
custodial symmetry naturally provides the quintessence to standard model singlet fermion 
couplings that sources neutrino masses. We also show that the model predicts a 
highly suppressed contribution to relativistic degrees of freedom from 
quintessential quanta at any late Universe epoch, ensuring the consistency of 
the model.
\end{abstract}

\maketitle
\section{Introduction}\label{sec:intro} %

Contemporary cosmological surveys 
\cite{Riess:1998cb,Perlmutter:1998np,Spergel:2003cb,Komatsu:2010fb,Ade:2015xua, 
Weinberg:2012es} have shown that the energy density of our Universe, in the 
framework of General Relativity, consists mainly of an unknown substance 
having the exotic property of overcoming the pull of gravity compelling our 
Universe to a stage of accelerated expansion. Whatever this component is made 
of 
it is known as dark energy (DE). The simplest candidate for DE is the 
cosmological constant ($\Lambda$) \cite{Weinberg:1996xe,Carroll:2000fy}, other 
more elaborated proposals invoke the existence of scalar fields 
\cite{Ratra:1987rm,Wetterich:1987fm,Frieman:1995pm} which near its vacuum state 
behave like $\Lambda$ and additionally have the advantage to allow a dynamics 
which could alleviate the problems of smallness and fine-tuning that 
$\Lambda$ has to deal with 
\cite{Weinberg:1988cp,Weinberg:2000yb,Bernard:2012nv,Wang:2017oiy}.

Since it was proposed, the feasibility of dynamic DE has been checked, as 
it can 
be seen for instance in early works like \cite{Coble:1996te}. To name just 
a more recent one see \cite{Tsujikawa:2012hv}. It is also expected to be checked in 
the near future through scheduled high precision probes like DESI 
\cite{Aghamousa:2016zmz}.

Several scalar fields have been proposed as DE, for instance \emph{Kessence} 
\cite{ArmendarizPicon:2000dh,ArmendarizPicon:2000ah}, \emph{Chaplygin Gas} 
\cite{Kamenshchik:2001cp,Bento:2002ps}, \emph{Phantom} 
\cite{Caldwell:1999ew,Carroll:2003st}, 
\emph{Hessence} \cite{Wei:2005fq,Wei:2005nw}, but among them likely the most 
known and studied 
is \emph{Quintessence} $(\Q)$ 
\cite{Caldwell:1997ii,Copeland:2006wr,Tsujikawa:2013fta}, which is 
thought as a canonical scalar field minimally coupled to gravity, its 
potential being flat enough to guarantee the slow-rolling
evolution of the field, which in turn is necessary to violate 
the strong energy condition and so to realize the accelerated cosmic 
expansion.

The cosmological evolution of $\Q$ has been studied widely regardless of its 
origin or the 
phenomenology of the high energy theory it could come from, to name only a few 
references see 
\cite{Harko:2013gha,Chiba:2009sj,Chiba:2009gg,Caldwell:2005tm}. It is possible 
to do this 
because to realize $\Q$ as DE it is only required the existence of a vacuum 
state that can be used as a classical source in Einstein's equations.

On the other hand, an underlying theory has to be considered when interactions 
between DE and 
other fields are taken into account, see for instance 
\cite{Anderson:1997un,Bean:2007ny} for DE 
and Dark Matter (DM) interactions, (for a review about DE and DM see 
\cite{Sahni:2004ai}). Other 
examples are the effects of coupling $\Q$ with ordinary matter, as it was 
revised in 
\cite{Carroll:1998zi}.
The first mention and a posterior study on the
possible connection among active neutrinos and $\Q$,
grounding the mass-varying neutrinos models, 
can be consulted in \cite{Fardon:2003eh,Peccei:2004sz}. A series of related 
studies can be found for instance in \cite{Afshordi:2005ym} and 
 \cite{MohseniSadjadi:2017pne, MohseniSadjadi:2018etb}.
The study of Yukawa couplings between DE and fermionic DM and the effects of radiative
corrections on the mass of $\Q$, as well as the proposal of multi-axion DE$/$DM
models and their cosmological evolution,  were addressed 
in~\cite{DAmico:2016jbm}. Early ideas regarding a possible connection among 
sterile Majorana neutrino masses and ultra-light bosons that could be $\Q$ were 
presented in~\cite{DAmico:2018hgc}, although no reference to any governing 
principle for that was given there.
To name only a few, studies on $\Q$ as an axionic particle or its connection 
with higher energy theories like string, superstring or M theory can be
found in \cite{Kumar:2013ecq,Albrecht:1999rm,Choi:1999xn}.

The dynamics of $\Q$ resembles that of the inflaton, which is the scalar field 
hypothesized in order to solve, among others, the horizon and flatness problems 
that non-inflationary (Friedmann) cosmologies suffered 
\cite{Guth:1980zm,Linde:1981mu,Linde:1983gd,Linde:1985ub,Linde:1986fd}.
Inflation assumes the early universe underwent an exponential expansion phase 
driven by a state of almost pure vacuum energy, that behaves like a cosmological 
constant,  generated through the slow-rolling evolution of the inflaton, but unlike
$\Q$, at a higher energy scale and 
totally dominating the content of the universe. Despite these facts, both 
dynamics are evidently similar to each other, and it seems  reasonable to assume 
that $\Q$ and the inflaton may be deeply interrelated.

Such is the line of thought of the cosmological unification idea  
presented in Ref.~\cite{PerezLorenzana:2007qv}. According to that, 
one can unify, in the field theory sense, using symmetries, both stages of 
accelerated expansion by relating inflation and 
quintessence fields with the degrees of freedom of a unique scalar field 
representation. In such an approach, DE would be just the remnant of the very 
early stages of cosmological evolution (see also Ref.~\cite{rosenfeld}).
Although the original model, based on the $SO(1,1)$ global symmetry, 
as discussed in \cite{PerezLorenzana:2007qv}, was  intended for phantom instead 
of $\Q$ as DE, on the basis of the same 
symmetry the Unification of inflation and $\Q$ is very well possible, as we will 
show below, by 
describing both the fields as associated to  the components of a doublet 
scalar representation. 

The interesting aftermath of this symmetry guided 
cosmological unification model, is that all possible interactions become very 
well defined at the Lagrangian level solely by the symmetry, in terms of a 
few field invariant couplings. That is the case of 
both scalar self-interactions as well as scalar to fermionic matter 
couplings. Hence, to the extent of a few fundamental parameters, all 
possible physics derived from the model becomes mostly determined.
Exploring these and probe up to what extent the $SO(1,1)$ cosmological 
unification model  can provide acceptable physical 
consequences is the main goal of the present paper.

Interestingly enough, as we will discuss later on, $SO(1,1)$ symmetry
does provide a set of bilinear field invariants that allow accommodating
inflation and quintessence dynamics from the most generic quadratic scalar
potential.
As explained briefly latter, more general potentials can be built by 
choosing higher-order invariants, nonetheless, we study the simplest one
as the first approximation to the phenomenology of our model, despite the fact
that the quadratic potential, in the inflation sector, is 
disfavoured by the Planck data \cite{Planck:2013jfk}.

To allow for a fermion to scalar coupling the symmetry enforces the
introduction of a fermion doublet and a singlet. We assume these fermions to be
right handed and singlets under the Standard Model (SM) of particle physics
symmetries and naturally identify them as neutrinos.
As expected, such Yukawa couplings would provide an inflaton decay channel for the
reheating after inflation. However, as we shall discuss, due to the symmetry,
the same set of couplings would keep right-handed neutrinos couple to the
quintessence field. The last would remain trapped in a false vacuum configuration
along the evolution of the observed Universe. According to quintessence model,
such a false vacuum is the actual source of the observed DE, yet, what becomes
even more interesting is the observation that in the context of our model,
this explanation of DE would also introduce a natural way to generate large masses
to right handed neutrinos, which would become connected to the cosmological
accelerated expansion. 

Right handed neutrino masses are the main known ingredient of the seesaw 
mechanism
~\cite{Minkowski:1977sc,GellMann:1980vs,Yanagida:1979as,Glashow:1979nm,Schechter:1980gr,Schechter:1981cv,Mohapatra:1980yp},
which provides a natural explanation to the tiny standard neutrino masses
observed in neutrino oscillation experiments (for a detailed discussion see \cite{Tanabashi:2018oca}), which are so far
bounded to be in the sub eV scale. (See also 
\cite{Vagnozzi:2017ovm,Vagnozzi:2018jhn} for  very strong
constraints on the sum of neutrino masses from cosmological data in the
context of both, constant and dynamical DE.)

In its simplest one family formulation a right handed singlet neutrino, 
$N$ is added to the SM particle content,
and the most general Lagrangian terms that  contribute to neutrino masses
are then written as
$y^\nu\bar L_\alpha \widetilde{H} N + (h.c) + M_{\mathsmaller{R}}\overline{N^{\mathcal C}}N$,
where $L$ stands for the SM lepton doublet, $H$ for the Higgs and $y^\nu$ for
the Yukawa couplings. By introducing the  Higgs vacuum, $\langle H\rangle$,
the first term becomes a Dirac mass term for the neutrino,
$m_{\mathsmaller{D}}\bar\nu N$,
where
 $ m_{\mathsmaller{D}} =  y^\nu\langle H\rangle/\sqrt 2~$,
which jointly to the 
Majorana mass term, provides a small effective mass for the standard 
neutrino that goes as
$m_\nu = (m_{\mathsmaller{D}})^2/M_{\mathsmaller{R}}~$.
Assuming an order one Yukawa, 
the only way to understand a sub eV $m_\nu$ is to have  $M_{\mathsmaller{R}}$  as 
large as $10^{13}$~GeV or so. Smaller values are yet possible if smaller Yukawa 
couplings are considered.
Nevertheless, notice  that Majorana mass enters as a free 
parameter in the theory, with no connection to the Higgs mechanism whatsoever.
Therefore, understanding neutrino masses with the 
seesaw mechanism becomes the search for an understanding of the origin of 
$M_{\mathsmaller{R}}$. Here is where the outcome of the $SO(1,1)$  model becomes of 
relevance by suggesting that such mass could actually have a cosmological 
origin, associated with the source of DE. 
This is a striking observation that deserves to be closely analyzed 
in order to establish its consistency in the cosmological setup
and doing so is the main goal of this paper.

To this end, we have organized our discussion as follows. In the next section, we 
introduce the Cosmological Unification model based in the SO(1,1) symmetry. 
There we present the Lagrangian of the model, which is based on the most 
general bilinear invariants built upon a dimension two fundamental 
representation to which cosmological scalar fields are assigned. We then 
discuss how inflation and quintessence emerge in the model. Right handed 
neutrinos are introduced to the model in section three. Yukawa couplings to the 
cosmological scalars are explored and the conditions upon which these get masses from 
the cosmic vacuum energy is discussed.
As his mechanism also implies that 
quintessence quanta, $\nchi$, can be excited in the primordial plasma from out 
of equilibrium right handed neutrino interactions, due to the same couplings that 
provide neutrino masses, in section four we explore the consequences of it, by 
studying the production of relativistic $\nchi$ fields through  
Boltzmann equations, which shows the consistency of the scenario with Big Bang 
Nucleosynthesis requirements.
Furthermore, since the seesaw mechanism 
also implies small $\nchi$ to active neutrino couplings, due to 
heavy to light neutrino mixings,  in section five we quantify the thermal 
corrections induced on quintessence mass due to cosmological neutrino 
backround, and check that, despite this, the slow-roll condition for $\nchi$ is 
always fulfilled, and hence the field keeps its DE behavior. 
Section six contains a short 
discussion and some final remarks  about our proposal. Finally, two 
appendices containing some technical details and relevant calculations are also 
included.

\section{The SO(1,1) cosmological model}

Following the motivations of the $SO(1,1)$ 
model as presented in Ref~\cite{PerezLorenzana:2007qv}, we consider the scalar  
doublet 
\begin{equation}\label{eq:scalar_doublet}
  \Phi = \doublet{\phi\\ \varphi},
\end{equation}
with $\phi$ and $\varphi$ complex scalar fields, which for convenience can be 
written in terms of four real fields as 
$$
\phi = \dfrac{1}{\sqrt 2} ( \phi_1 + i\phi_2), \qquad \varphi = \dfrac{1}{\sqrt 
2} (\varphi_1 + i\varphi_2).
$$
 This representation
transforms under the global $SO(1,1)$ group as
$\Phi \longrightarrow g_\alpha\Phi$, 
where $g_\alpha$ stands for an arbitrary element in the corresponding 
$SO(1,1)$ matrix representation, whose exponential mapping is in general given 
by
\begin{equation}\label{eq:SO_1_1_element}
g_\alpha = e^{i\alpha\sigma_1}, \quad\alpha \in \Reals, 
\end{equation}
with $\sigma_1$ the first Pauli matrix. 

There are four bilinear invariants formed with this doublet 
\cite{PerezLorenzana:2007qv}:
\begin{equation}\label{eq:the_scalar_invariants}
\begin{split}
  &\Phi^\dagger\Phi  = |\phi|^2 + |\varphi|^2,\qquad
  \Phi^\dagger\sigma_1\Phi  =  \phi^* \varphi + \varphi^* \phi,\\ 
  &\Phi^{\mathsmaller T}i\sigma_2\Phi  =  \phi\varphi - \varphi\phi,\hspace{2.9ex} 
  \Phi^{\mathsmaller T}\sigma_3\Phi  =  \phi^2 - \varphi^2,
\end{split}
\end{equation}
with $ \sigma_2$ and  $\sigma_3,$ the other two Pauli matrices. 
Clearly, the kinetic term $\partial_\mu\Phi^\dagger\partial^\mu\Phi$ belongs to 
the first class of invariants in the above equation. The potential of the 
model, on the other hand, is restricted  to be built out of 
these invariants in order to keep the symmetry. 

It is worth
noticing that these terms still allow for some diversity on the 
possible cosmological potentials one may consider. In the case of real field 
representations, for instance,  first and third invariants can 
be added together to provide for a whole class of systems where the fields 
have an independent evolution, simply because one can write 
$\phi^2 = \Phi^\dagger\Phi + \Phi^{\mathsmaller{T}}\sigma_3\Phi$, and 
$\varphi^2 = \Phi^\dagger\Phi - \Phi^{\mathsmaller{T}}\sigma_3\Phi$. In such a case, the 
potentials $U(\phi^2)$ and $V(\varphi^2)$ written in terms of such combinations 
would always have a quadratic dependence on the fields. Of course, such a 
scenario implies the removal of the $\Phi^\dagger\sigma_1\Phi$ term from the 
theory, but as stated in Ref.~\cite{PerezLorenzana:2007qv} this could be done 
by 
noticing that such a term is actually a pseudoscalar bilinear under 
the parity transformation defined as $\Phi\rightarrow\sigma_3\Phi$, which can 
easily be added to the model. 
Such a construction, however, ignores the most general complex nature of 
the cosmological field $\Phi$ and we will avoid it.

Next, for our model, we  consider the most general 
theory we can build out of the invariant terms in 
 Eq.~(\ref{eq:the_scalar_invariants}), but considering for simplicity only mass 
like terms in the potential.
As it should be clear, more general potentials based on these same bilinears 
are also possible, but considering this simplest form, although disfavoured by 
Planck data,  will suffice for our propose. 
Therefore, the Lagrangian we consider would be
\begin{equation}\label{eq:scalar_sector}
\Lag_{\Phi} = \partial^\mu\Phi^\dagger\partial_\mu\Phi - V(\Phi),
\end{equation}
where the potential is formed from the most general linear combination of the 
non trivial invariants,
\begin{equation}\label{eq:potential_Phi}
V(\Phi) = \Phi^\dagger\left(\alpha_0\mathbb I + \alpha_1\sigma_1\right)\Phi  
+ \alpha_3\Phi^{\mathsmaller T}\sigma_3\Phi + h.c.
\end{equation}
Here  $\alpha_{i=0,1,3}$ are mass dimension two quantities which in general 
can be complex. As the model intends to incorporate inflation, we should assume 
that the involved scales are naturally large, perhaps as few orders below the 
Planck scale, $m_{\text{pl}}$. Also, a contraction with the background 
Friedmann-Robertson-Walker 
metric should be understood in the kinetic terms. In order to identify the 
dynamics of the so constructed cosmological model, we need to explore the 
potential in detail and identify the proper set of initial conditions that 
should give rise to inflation and DE.

As explained in detail in Appendix A, the above generic potential can be 
diagonalized using an orthogonal rotation, $\mathbb S^\prime$,  on the four 
dimensional field space of initial real field components, such that we can use 
the new fields defined as
$(\Q_1 ,  \Q_2, \xi_1 , \xi_2 )^{\mathsmaller{T}} = {\mathbb S^\prime} 
 ( \phi_1, \phi_2, \varphi_1,\varphi_2)^{\mathsmaller{T}}$, to build the mass eigenstate 
complex scalars 
\begin{equation}
\Q = \dfrac{1}{\sqrt 2}(\Q_1 + i\Q_2), \qquad \text{and} \qquad \xi = 
\dfrac{1}{\sqrt 2}(\xi_1 + i\xi_2), 
\end{equation}
out of which the Lagrangian simply becomes [see Eq.~(\ref{eq:Lag_X})]
\begin{eqnarray}\label{QxiModel}
\Lag_\Phi 
&=& \partial^\mu\rbvarphi^\dagger\partial_\mu\rbvarphi - 
\rbvarphi^\dagger\mathbb M \rbvarphi~,\\
&=& 
\partial^\mu \Q^\dagger\partial_\mu \Q +
\partial^\mu\xi^\dagger\partial_\mu\xi - m^2|\Q|^2 - M^2|\xi|^2~,
\end{eqnarray}
where 
\begin{equation}  
\rbvarphi =
  \begin{pmatrix}
    \Q \\ \xi    
  \end{pmatrix}~ \quad\text{and}\quad 
  \mathbb M =
\begin{pmatrix}
m^2 & 0 \\  0 & M^2
\end{pmatrix}~.
\end{equation}
As stated in Appendix A, above masses, written in terms of $\alpha_i$, are 
expressed as
\begin{equation}
 M^2= \mu_0^2 + \mu^2~; \quad\text{whereas}\quad m^2 = \mu_0^2 - \mu^2~;
\end{equation}
where $\mu_0^2 = 2Re~\alpha_0$ and 
$\mu^2 = 2\sqrt{(Re~\alpha_1)^2 + |\alpha_3|^2}$~.

Notice that even though in  Eq.~(\ref{eq:scalar_sector}) we started with a 
coupled 
system of complex fields, after field rotation  
we have ended with a new description where  $\Q$ and 
$\xi$ degrees of freedom had been decoupled. 
However, we should also notice that, even though this is a more 
suitable way of writing the potential, it is on the cost of hiding the 
$SO(1,1)$ symmetry, which now is not explicit in the Lagrangian.

Furthermore, the  potential in  Eq.~(\ref{QxiModel}) shows no explicit 
dependence on the phase fields which suggests that they should not play any 
fundamental role in the slow-roll evolution phase of the background 
cosmological system. In accordance with this, for purposes of simplicity, we 
shall proceed with the analysis of the  cosmological model by only 
considering the modular field components as a good first approximation, fixing 
the phases to zero. However, as one may still consider worth 
asking about the role played by these field phases on other effects of 
cosmological interest, particularly as in the DE sector where  this phase could 
play a regulatory role, as it is done, for instance, in spintessence 
models~\cite{Caldwell:2009ix,Boyle:2001du}, we are addressing the issue in some 
detail in Appendix B, where it is shown that the system dynamics of the 
background universe does indicate that it is indeed consistent to choose the 
initial phase value being zero, such that the phase does not 
evolve.

The supplementary condition $\mu_0^2 \approx \mu^2$, which is consistent with  
the assumption that all 
involved scales were naturally about the same order,  allows incorporating
a fine-tuning in the masses just to have $M^2\gg m^2$, which 
permits to identify $\xi$ as the inflation field, and $\Q$ as the quintessence 
source of DE. As a matter of fact, in such a case the cosmological 
system involves the independent evolution of two fields that fall on a 
paraboloidal potential from some given initial condition towards the absolute 
minimum located in $\xi=\Q=0$. Clearly, for $M^2\gg m^2$, the potential is 
steeper along $\xi$ direction with $\Q$ behaving almost like a flat direction. 
Assuming that the initial condition is such that $\langle\xi\rangle\sim 
\langle \Q\rangle\sim m_{\text{pl}}$, 
in the slow-roll regime the source for inflation in the model would then be  
proportional to the squared modulus of the inflaton, as it is done in chaotic 
inflation. Similarly, the source for DE is proportional to 
the squared modulus of $\langle \Q\rangle$. 
According to the standard dynamics,
$\xi$ should slow-roll down the potential 
towards the local minimum at $\xi=0$ but where $\Q$ is frozen at its initial 
value, $\langle \Q\rangle$ due to its 
small mass since $m\ll \mathsf H\approx M$, with $\mathsf H$ the Hubble 
parameter, and thus $\dot \Q/\Q\approx m^2/\mathsf H\approx 0$, until $\mathsf 
H$ catches with $m$ scale. Effectively, $\Q$ would behave most of the time 
as a perfect fluid with an equation of state $p=\omega \rho$ with 
$\omega = -1$. Eventually, $\xi$ would exit inflation and 
suddenly evaporates and reheats the Universe. 
As usual for chaotic inflation, the observed amount of density perturbations in
the cosmic microwave background would require $M\approx 10^{-5} m_{\text{pl}}$. 
$\Q$, on the other hand,  should stay fixed at its initial value
along most eras of evolution, until matter density, $\rho_m$,  catches 
with quintessence false vacuum energy density, 
\begin{equation}\label{eq:false_vacuum_energy_density}
\rho_{\langle \Q\rangle} = \dfrac{1}{2}m^2\langle \Q \rangle^2~,
\end{equation}
near the coincidence era. After that, $\Q$ gets 
released and starts slow-rolling down towards its true minimum at zero. 
Most  of the $\Q$-models use this expression to rewrite the observed  DE 
density \cite{Tanabashi:2018oca}, 
$\rho_{\mathsmaller{DE}} = M_{pl}^2\Lambda\approx 2.53 \times 
10^{-47}$~GeV$^4$, where $M_{\text{pl}} = m_{\text{pl}}/\sqrt{8\pi}$ is the 
reduced Planck mass,
such that
$$
m^2 = 2\dfrac{M_{pl}^2}{\langle \Q\rangle^2}\Lambda~.
$$
Therefore, with
$\langle \Q\rangle \sim m_{\text{pl}}~$,
the mass of $\Q$ should be 
as small as
\begin{equation}\label{eq:scalar_mass_required}
m\approx 5.8\times 10^{-34}~\text{eV}~,
\end{equation}
to provide a successful scenario.
The smallness of this parameter indicates
the need for a fine-tuning as large as in the cosmological constant problem.

We would like to finish this section by making some comments about the smallness 
of this scale.
It has been noticed that the tiny mass of $\Q$  is unstable under 
radiative corrections due to quadratic divergences, in such a way that the required flatness of the potential could be wiped and so, the slow-roll condition \cite{Kolda:1998wq}.  This is a generic illness of any 
interacting scalar field theory, commonly known as the hierarchy problem.
In order to keep the physical  
mass around the required order, it is necessary to introduce 
further large  fine-tunings on all loop order corrections.

This problem is commonly overcome in the context of supersymmetry, where 
quadratic divergences are exactly canceled by  
superpartner contributions, and hence, the mass is kept under control.
Such is the case, for instance,  of the Higgs mass in the context of  
Supersymmetric Standard Model extension.
Nevertheless, in the case of $\Q$, it has been observed that simply adding 
supersymmetry might not solve the whole problem, since remaining 
corrections due to supersymmetry breaking might still be large (see 
\cite{Kolda:1998wq,Hall:2005xb}).
Additionally, Q-mass could be stabilized by invoking Goldston 
symmetries,  where  Q is assumed to be a pseudo-Goldstone boson belonging to a 
higher dimensional space in which the supersymmetric breaking scales are 
suppressed, thus, this boson appears as an effective boson that preserves its 
stability in four dimensions  (see for instance 
\cite{Tsujikawa:2013fta,Hall:2005xb,Burgess:2003bia}).  

Whether our model could be extended to consider supersymmetry or be embedded in a more fundamental theory (perhaps SUGRA or even String Theory), where $\xi$ and $\Q$ fields arise as effective degrees of freedom, such that the large hierarchy problem gets under control,  is an issue we still have to explore, so it remains as potentially troublesome for the model, as it is for most quintessence models.

\section{Adding fermions: Reheating and neutrino mass}

Reheating after inflation in the usual approach uses the sudden decay of the 
inflaton into other particles in order to inject matter in an otherwise 
empty Universe. In accordance with the global $SO(1,1)$ 
symmetry we adopted as the protective one for our cosmological model, 
the minimal fermionic matter content is accounted by 
introducing a total of 
three spinorial  fields, $N_{i=0,1,2}^{\dot a}$,  
two of then arranged into a doublet  
\begin{equation}\label{eq:neutrino_doublet}
  \Psi =
  \doublet{   N_1^{\dot a}\\N_2^{\dot a} },
\end{equation}
and the remaining one treated as a singlet. 
We choose fermions to be (two-component) right handed Weyl fields,  such that 
they can be identified with those
usually  introduced in extensions of the standard model of 
particle physics in order to have massive neutrinos through the seesaw 
mechanism. Thus, the two-component spinorial index $\dot a = 1,2$ 
and Dirac matrices are written as
\begin{equation}\label{eq:Dirac_matrices}
\gamma^\mu =
\begin{pmatrix}
  0 & \sigma_{\ a\dot c}^\mu \\
  \bar\sigma^{\mu \dot a c} & 0
\end{pmatrix},
\end{equation}
with,
$$
\sigma^\mu_{\ a\dot a} = \left(\mathbb I,\  \mathbf 
\sigma\right),\quad\bar\sigma^{\mu\dot a a} = \left( \mathbb I,\  
-\mathbf\sigma\right),\quad \bar\sigma^{\mu\dot a a} = 
\epsilon^{\dot a\dot b}\epsilon^{ab}\sigma^\mu_{\ b\dot b},
$$
and  $\mathbf \sigma = \left( \sigma_1,\ \sigma_2,\ \sigma_3  \right)$. In this
notation, the charge conjugation matrix and the $\beta$ matrix (which is
numerically equal to $\gamma^0$ but carrying different index structure), are
respectively given by
\begin{equation}\label{eq:C_and_beta_matrices}
C =
\begin{pmatrix}
  \epsilon_{ac} & 0 \\
  0 & \epsilon^{\dot a \dot c}
\end{pmatrix},
\quad
\beta =
\begin{pmatrix}
  0 & \delta^{\dot a}_{\ \dot c}\\
 \delta_a^{\ c} & 0 
\end{pmatrix}.
\end{equation}

From each Weyl field,  a four-component ($a,\dot a = 1,2$) sterile Majorana neutrino  is
built by writing
\begin{equation}\label{eq:Majorana_sterile_neutrino}
\psi_i =
\doublet{N_{i a}^\dagger\\ N_i^{\dot a}},
\end{equation}
where $N_{ia}^\dagger$ is the charge conjugate of the right handed Weyl field, 
given by
$N_{ia}^\dagger =
  (N_i^{\dot a})^{{\mathcal C}}$.
The previous can be seen from  Eq.~(\ref{eq:Majorana_sterile_neutrino}) and 
$\psi^{\mathcal
C} = C\bar\psi^{\mathsmaller T}$ with the application of
 Eq.~(\ref{eq:C_and_beta_matrices}). 
The doublet in  Eq.~(\ref{eq:neutrino_doublet}) transforms under  $g_\alpha \in 
SO(1,1)$ as
\begin{equation}\label{eq:spinor_rotation}
  \doublet{
    N_1^{\dot a} \\ N_2^{\dot a}
  }
  \overset{g_\alpha}{\longrightarrow}
  e^{i\alpha\sigma_1}
  \doublet{
    N_1^{\dot a}\\N_2^{\dot a}
  } =
  \doublet{
    N_1^{'\dot a} \\ N_2^{'\dot a}
  },
\end{equation}
with the new Weyl fields arising from combinations and global phase changes of
the previous ones. It is important to note that since the Weyl fields admit
global phase transformations, it will be always possible to build a new
four-component sterile Majorana neutrino
$$
\psi'_i =
\doublet{
  N_{i a}'^{\dagger} \\ N_i'^{\dot a}
},
\quad
\text{
such that
}
\quad
\psi_i = \psi_i^{{\mathcal C}} \overset{g_\alpha}{\longrightarrow}
 \psi'_i = \psi_i'^{\mathcal C},  
$$
therefore the  transformation of the field $\psi_i$ induced by the 
$SO(1,1)$ rotation
in  Eq.~(\ref{eq:spinor_rotation}) does not violate the Majorana condition. 

With these conventions, the general fermion kinetic terms for the Majorana fields
become
$
\dfrac{1}{2}\bar\psi_i i \gamma^\mu\partial_\mu\psi_i = 
N_i^{\dagger a} i \sigma_{a\dot c}^\mu\partial_\mu N_i^{\dot c},
$
where a  background metric contraction should be understood as before. 
Next, it is easy to see that one can write the kinetic terms in a clearly 
$SO(1,1)$ and Lorentz invariant form, as
\begin{equation}\label{eq:fermion_kinetic_terms}
\Lag_{\Psi} = N_0^{\dagger a} i \sigma^\mu_{a\dot c}\partial_\mu N_0^{\dot c} + 
\Psi^\dagger i \sigma^\mu\partial_\mu\Psi~.
\end{equation}
On the other hand, by taking the Hermitian conjugate of $N_0^{\dot a}$
and the fermion and scalar doublets, the most general Yukawa interaction terms 
from the linear combination of the invariants one can build are
\begin{multline}\label{eq:interaction_Phi}
-\Lag_I = N_{0\dot a} \big\{ a_0\Phi^\dagger\Psi + 
a_1\Phi^\dagger\sigma_1\Psi \\ + a_2\Phi^\mathsmaller{T}i\sigma_2\Psi + 
a_3\Phi^\mathsmaller{T}\sigma_3\Psi\big\} + h.c.,
\end{multline}
where $a_{i=0,\dots,3}$ are complex dimensionless couplings.

Notice that analogous to the invariant terms which appear in
 Eq.~(\ref{eq:the_scalar_invariants}), there exist bilinear $SO(1,1)$ 
invariants that are formed from the fermion doublet taken with itself,
$\Psi^\dagger\Psi,\quad\Psi^\dagger\sigma_1\Psi,\quad\Psi^\mathsmaller{T}i
\sigma_2\Psi\quad \text{and}\quad\Psi^\mathsmaller{T}\sigma_3\Psi,
$
which,  however, are not Lorentz
invariant objects and therefore we take them off from the Lagrangian.

It is also worth asking if there are allowed mass terms for the fermions. We
note that such terms can be built by defining an additional doublet formed
from the charge conjugate fields of $N_{i=1,2}^{\dot a}$, as
\begin{equation}\label{eq:charge_conjugated_neutrino_doublet}
  \Psi^\mathcal{C} =
  \doublet{
    N_{1a}^\dagger \\ N_{2a}^\dagger
  }.
\end{equation}
The following product, which is a Lorentz-invariant scalar
\begin{equation}\label{eq:forbidden_mass_term}
\Psi^{\mathcal{C}\dagger}\Psi + h.c. = N_{1\dot a}N_1^{\dot a} + 
N_{2\dot a}N_2^{\dot a} + h.c., 
\end{equation}
clearly produces Majorana mass terms 
$\left(\psi_i^{\mathsmaller T}C^\dagger\psi_i\right)$ for the fields 
$\psi_{i=1,2}$, however, in order to
get a consistent transformation of 
$\Psi^{\mathcal C}$ under the symmetry, it is necessary to 
impose the  condition that
$$
N'^{\dagger}_{ia} = (N'^{\dot a}_i)^{\mathcal C},
$$
which means that the components of the charge conjugate rotated doublet
$\Psi'^{\mathcal C}$ must to be equal to the charge conjugate components of the
rotated doublet $\Psi'$. In order to achieve this, the doublet
in  Eq.~(\ref{eq:charge_conjugated_neutrino_doublet}) has to transform with the
Hermitian conjugate matrix $g_\alpha^\dagger$, as can be checked by means of the
two-dimensional matrix representations. 
Consequently, the term in  Eq.~(\ref{eq:forbidden_mass_term}) is not invariant 
under $SO(1,1)$ rotations and we must remove it from the Lagrangian. The same 
occurs for all
the terms formed from  Eq.~(\ref{eq:charge_conjugated_neutrino_doublet}) and
(\ref{eq:neutrino_doublet}). On the other hand, a mass term for $N_0^{\dot a}$
does is allowed by the $SO(1,1)$ symmetry because it transforms as a singlet,
however, we note that the interaction sector in  Eq.~(\ref{eq:interaction_Phi}) 
is invariant under the following $U(1)$  transformation
\begin{equation}\label{eq:U_1_transformation}
\Psi \longrightarrow e^{iq}\Psi, \qquad N_0^{\dot a} \longrightarrow 
e^{iq_0}N_0^{\dot a}, 
\end{equation}
as long as $q = -q_0$. So, the fields $N_{i=1,2}^{\dot a}$ transform with
the same charge and $N_0^{\dot a}$ does it with the opposite. Thereby, by
imposing invariance under $U(1)$ in the fermion sector, which implies lepton
number conservation, we remove the singlet's mass term. We note that the same
argument can be invoked in order to forbid mass terms for the fermions
$\psi_{i=1,2}$, but this only confirms what the $SO(1,1)$ symmetry suggests.

Finally, the complete Lagrangian we are left with, is
\begin{equation}\label{eq:complete_lagrangian}
\Lag = \Lag_{\Phi} + \Lag_{\Psi} + \Lag_{I},
\end{equation}
where the three sectors are respectively given by  Eq.~(\ref{eq:scalar_sector}),
 Eq.~(\ref{eq:fermion_kinetic_terms}) and  Eq.~(\ref{eq:interaction_Phi}).
The above Lagrangian is the most general one that can be written with $SO(1,1)$
bilinear invariant terms, it is also Lorentz invariant, $P$ (as long as both
scalar fields transform with the same parity phase) and $CP$ invariant. As
mentioned above the fermionic sector is $U(1)$ invariant, similarly, there is
$U(1)$ invariance in the scalar sector, as long as both $\phi$ and $\varphi$
transform with the same charge.

\subsection{Reheating}

By performing the rotation in field space that diagonalizes the scalar sector 
and allows to identify the inflaton and quintessence fields, one has also to 
redefine the general Yukawa couplings introduced in 
 Eq.~(\ref{eq:interaction_Phi}). After some algebra, as explained in 
detail in  Appendix A, scalar to fermion couplings [see  
Eq.~(\ref{eq:interaction_F})] can be put into the following simple expression
\begin{equation}\label{eq:interaction_F_ppal}
-\Lag_I = N_{0\dot a}\{ \rbvarphi^\dagger\mathbb G_1\mathbf F + 
\rbvarphi^\mathsmaller{T}\mathbb G_2\mathbf F \} + h.c.,
\end{equation}
where the new coupling constants, which are just simple linear combinations of 
the original $a_i$ constants written in Eq.~(\ref{eq:interaction_Phi}),  are 
contained in the matrices [see Eq.~(\ref{eq:couplings_matrices_app})]
$$
\mathbb G_1 =
\begin{pmatrix}
0 & g_2 \\ h_1 & 0
\end{pmatrix}
\qquad
\mathbb G_2 =
\begin{pmatrix}
g_1 & 0 \\ 0 & -h_2
\end{pmatrix}.
$$
Here, the Weyl fields $F_{i=1,2}^{\dot a}$  are the components 
of the doublet
\begin{equation}
  \mathbf F =
  \doublet{
    F_1^{\dot a} \\ F_2^{\dot a}
  },
\end{equation}
which arises from Eq.~(\ref{eq:neutrino_doublet}) after performing a 
$SO(2)$ rotation, $e^{-i\sigma_2 \pi/4}\Psi = \mathbf F$, 
as can be seen in equation~(\ref{eq:Psi_to_F}). 
Notice that this rotation also transforms the spinor kinetic terms, which 
remain diagonal [see equation (\ref{eq:fermion_kinetic_terms_2})].
Clearly, as for the scalar sector, after the transformations  the $SO(1,1)$ 
symmetry is not explicit
in the Yukawa Lagrangian anymore. The assumed  $U(1)$ 
symmetry imposed in the fermion sector remains explicit, on the other hand.

Former couplings can be written in a more useful way, as
\begin{multline}\label{Yukawas}
-\Lag_I = N_{0\dot a}\{ g_1\Q F_1^{\dot a} + g_2\Q^* F_2^{\dot a} \\ + h_1\xi^* 
F_1^{\dot a} - h_2\xi F_2^{\dot a}  \} + h.c.
\end{multline}
The last two terms of Eq.~(\ref{Yukawas}) provide the inflaton decay channels, 
$\xi\rightarrow N_0 F_i$, that are required for reheating after inflation. 
The sudden evaporation of inflaton energy would inject entropy to the emptied 
Universe by inflation. Assuming that such a process is 
efficient enough, the reheating temperature  should be 
$T_r\sim 6\times 10^{-3}~max\{|h_1|,|h_2|\}~M_{pl}$.
Since the fermions on final states are assumed to be 
right handed neutrinos they should provide the portal, through the standard 
couplings $\bar{L}\widetilde{H} N_0$ and $\bar{L}\widetilde{H} F_i$, to produce 
all types of SM fields, which in turn should 
thermalize producing the primordial plasma. 

\subsection{SO(1,1) as a flavor symmetry?}

Notice that the $\bar{L}\widetilde{H} F_i$ couplings would explicitly violate 
the $SO(1,1)$ symmetry unless the SM matter fields had non-trivial 
transformations. Of course, we can proceed with our study by assuming this, in 
which case, $SO(1,1)$  would be a symmetry of the cosmological and sterile 
sectors only, which is broken by the Yukawa couplings to the standard fields.
Such a scenario mimics the construction of models used to explain supersymmetry 
breaking, where two sectors exist in the theory, one called visible, where the 
symmetry is respected, and another where it is violated, called the hidden 
sector. A messenger, that couples both the sectors, communicates the breaking of 
the symmetry on the second to the first sector, producing a rich phenomenology 
and even dark matter candidates~\cite{hidden} have been considered in such 
models. Here, on the contrary, the hidden sector (cosmological plus right-handed 
neutrinos) has the symmetry, whereas the  SM sector does not transform under 
$SO(1,1)$. Thus, the coupling among them would be the source of the breaking and 
the right-handed neutrinos themselves should be the messengers to carry this 
explicit breaking to the cosmological sector. Consequences of this mechanism 
would appear through loop quantum corrections, and so we do not expect them to 
be as relevant as to affect the classical configuration of the quintessential 
field, nor the physics we are discussing below. Yet, this is an issue that may 
deserve further analysis.

The alternative, on the other hand, is quite interesting. Global flavor 
symmetries have been longly considered as a way to understand the fermion mass 
spectrum as well as flavor mixings appearing in charged weak interactions. In 
the lepton sector, such mixings are responsible for neutrino oscillation 
phenomena. Extending the cosmological symmetry to involve also SM fields drops 
right in this class of symmetries. Indeed, it turns out that it is possible to 
imagine a simple model extension where  $SO(1,1)$ is promoted to be a global 
flavor symmetry for the leptonic SM sector, at least.  

To be specific, let us consider the following lepton matter content assignments 
under $SO(1,1)$. We assume both, left and right-handed $\mu$ and $\tau$ type 
flavors  as belonging to doublets, written as $$
2_L = \doublet{L_\mu\\L_\tau}\quad\text{and}\quad 2_R = \doublet{\mu_{\mathsmaller R}\\\tau_{\mathsmaller R}}.
$$
Electron type fermions, $L_e$  and $e_{\mathsmaller R}$, shall belong to 
singlets. Right-handed neutrinos, of course, would be given by those 
representations introduced in our cosmological model above. Furthermore,  we 
take an extended Higgs sector, formed by the standard $SO(1,1)$ singlet Higgs, 
$H$, and a doublet of Higgses, written as
$$
2_H = \doublet{H_1\\H_2}.
$$
With these matter content, we write the most general Yukawa couplings given by 
\begin{equation}
 \bar{2}_L\cdot 2_H e_R + 
 \bar{2}_L\cdot 2_R H + 
 \bar{L}_e 2_H\cdot 2_R + 
 \bar{L}_e H e_R +
 h.c. 
\end{equation}
and
\begin{equation}
\bar{2}_L\cdot\widetilde{2}_H N_0 +
\bar{2}_L\cdot\Psi\widetilde{H} +
\bar{L}_e\widetilde{2}_H\cdot \Psi + 
\bar{L}_e \widetilde{H} N_0  +
h.c.~,
\label{Dmass}
\end{equation}
where, in each term, the contribution of all $SO(1,1)$ relevant invariants should 
be understood.
For instance, for the two fermion doublet couplings, we should read 
\begin{align*}
\bar{2}_L\cdot 2_R &= 
2_L^\dagger
\gamma^0
(f_1+f_2\sigma_1)
                     2_R\\
                   &=
                     \doublet{\bar L_\mu\ \bar L_\tau}
                     (f_1+f_2\sigma_1)\doublet{\mu_{\mathsmaller R}\\\tau_{\mathsmaller R}}
~, 
\end{align*}
where $f_{1,2}$ are Yukawa couplings.
Clearly, this invariant contains bilinears of the same type as the first and 
second of those shown in (\ref{eq:the_scalar_invariants}). Similar expressions 
should be written for
$\bar{2}_L\cdot\Psi$,  $\bar{2}_L\cdot 2_H$ and 
$\widetilde{2}_H\cdot\Psi$. 
For the other bilinears, like those of the third and fourth type shown on 
(\ref{eq:the_scalar_invariants}), we get invariants such as \begin{align*}
  2_H\cdot 2_R &= 2_H^{\mathsmaller T}(ik_1\sigma_2 + k_2\sigma_3)2_R,\\
  &=  \doublet{ H_1\ H_2 }(ik_1\sigma_2 + k_2\sigma_3)\doublet{\mu_{\mathsmaller R}\\\tau_{\mathsmaller R}},
\end{align*}
with $k_{1,2}$ as the Yukawa couplings.
Same type of couplings should be 
understood from  $\bar{2}_L\cdot\widetilde{2}_H$.
Note that all terms in Eq.~(\ref{Dmass}) are exactly of the 
$\bar{L}\widetilde{H} F_i$ type, as required for a successful reheating process 
in our cosmological model, through the Higgs portal.

After spontaneous symmetry breaking,  the above terms would provide the 
nine mass terms of the charged lepton mass matrix, $M_\ell$, as well as the nine  
Dirac neutrino mass terms that will contribute to the seesaw mechanism.
Without further assumptions, all such terms should be expected to be non zero. 
Additional flavor symmetries might be introduced in order to generate specific 
textures.
After diagonalizing both the sectors, flavor mixings would appear in the 
charged weak interactions given by the mixing matrix 
$U_{mix} = V_\ell^\dagger U_\nu $, where $U_\nu$ ($V_\ell$) denotes the 
rotation matrix of the left-handed neutrino (charged lepton) sector used in 
the diagonalization process, as it is well known. 

As the interest in the present work is to analyze the cosmological setup, we 
will not further discuss the details and the phenomenology that should arise 
from this flavor model. An extended analysis of this will be presented 
elsewhere. Nevertheless, what we have discussed above does serve to show that 
$SO(1,1)$ symmetry may also work as a SM flavor symmetry, allowing enough 
freedom to accommodate lepton masses and mixings,  at the cost, of course, of 
extending the Higgs sector. The extra Higgses would have little impact on what 
follows, and thus we will proceed with our discussion without considering them 
explicitly.

\subsection{Sourcing neutrino mass with DE}

At the end of inflation the  $\xi$ field evaporates completely, 
such that its energy density becomes null, sitting the inflaton field at its 
zero value which makes its couplings of no further relevance for thermal 
history. 
On the other hand, as we have already discussed in the previous section, the $\Q$ 
field would remain trapped on its initial homogeneous configuration all along 
the Universe evolution, perhaps changing quite slowly until recent times,
when it is still slow-rolling down its almost flat potential while causing 
the Universe accelerated expansion. 

By inserting the $\Q$ false vacuum, conveniently defined as 
$\langle \Q\rangle/\sqrt{2}$, back in Eq.~(\ref{Yukawas}), one immediately 
realizes that due to the couplings provided by the $SO(1,1)$ model, DE naturally 
generates masses for the right handed neutrinos, given as
\begin{equation}
\label{N_mass}
\Lag_{m} =  m_1 N_{0\dot a} F_1^{\dot a} + m_2 N_{0\dot a} 
F_2^{\dot a} + 
h.c.~,
\end{equation}
where $m_i = g_i \langle \Q\rangle/\sqrt{2}$. 
These mass terms, as discussed in detail in 
Appendix B, give rise to two degenerate massive 
Majorana neutrinos, $\nu_{1,2}$, for which one can write
\begin{equation}\label{eq:Lag_mk_equiv}
-\Lag_m = \dfrac{1}{2}m_k\left(\bar\nu_1\nu_1 + \bar\nu_2\nu_2\right)~. 
\end{equation}
This is a striking result, which  connects the seesaw mechanism, 
and thus the origin of standard neutrino mass, to the origin of DE.

Here, we have implicitly written the
Majorana condition, namely $\bar\nu = \nu^{\mathsmaller{T}}C$, with $C$ the
charge conjugation matrix [see equation (\ref{eq:C_and_beta_matrices})].
Likewise, the
mass $m_k$ appearing in Eq.~(\ref{eq:Lag_mk_equiv}), as defined in
(\ref{eq:neutrino_mass_def_app}), is given by
\begin{equation}\label{eq:neutrino_mass_def_ppal}
m_k = \dfrac{a_c}{\sqrt 2}\langle \Q\rangle~,
\end{equation}
where  the effective coupling $a_c=\sqrt{|g_1|^2 + |g_2|^2}$~.
We note that by choosing $a_c$  in the 
interval 
$10^{-5} \lesssim a_c 
\lesssim 10^{-3}$, which seems reasonable,
we can get right handed neutrino masses in the range of  $10^{14}~\text{GeV} \lesssim 
m_k \lesssim 10^{16}~\text{GeV}$, which are values 
around those needed to implement the standard seesaw mechanism.

Another immediate outcome of the present model is the alignment to mass terms 
of couplings among quintessence quantum excitations, $\nchi$, and neutrinos.
Setting in the excitations over the false vacuum, 
by redefining $\Q=(\langle \Q\rangle + \nchi)/\sqrt{2}$, it is clear that after 
diagonalizing fermion masses, one gets 
\begin{equation}\label{eq:Lag_IQ_equiv}
-\Lag_{I \nchi} = \dfrac{a_c}{2\sqrt 2}\nchi\left( \bar\nu_1\nu_1 + 
\bar\nu_2\nu_2\right).
\end{equation}
This coupling  has relevance for thermal history. 
Equations (\ref{eq:Lag_mk_equiv}) and (\ref{eq:Lag_IQ_equiv}) show that
only two neutrinos are massive and interact with the DE field. The third
neutrino remains massless and decoupled.
Heavy neutrinos will eventually become non-relativistic in the very early 
stages of the Universe and decay. Main decay process would go into SM particles as 
$\nu_i\rightarrow L H$, injecting entropy to the primordial plasma.
However, there could be an increase in the relativistic energy density due to out-of-equilibrium
processes allowed by (\ref{eq:Lag_IQ_equiv}),
since quintessence is a rather ultralight field, and the co-annihilation process
$\nu\nu\rightarrow \nchi\nchi$ will populate this  degree of 
freedom as we will examine in the next section.

Notice that once seesaw mechanism gets introduced, after SM  symmetry 
breaking,  this will produce a mixing among the heavy and the 
standard neutrino states, of order $m_{\mathsmaller D}/m_k$. Such a mixing 
would in turn introduce an effective coupling among  $\nchi$ and the 
light active neutrinos, which, together with the photons of the CMB, permeate 
the Universe in the form of radiation. As a consequence, a thermal 
mass correction to $\nchi$ potential,  due to cosmological neutrino 
background, had to be taken into account since  it could eventually overcome 
the Hubble parameter, breaking the slow-roll condition. We will also address 
this issue later on (see  section five).

It is worth mentioning that although the above analysis  assumed 
neglecting phases for the fields, their inclusion has little impact on 
our main conclusions. To state our point we are including in Appendix B a
detailed discussion of the changes and effects that are involved when the phase 
of the scalar fields are considered. In particular, we notice that the phase of 
the scalar DE field does not take part in the
interaction sector  beyond the term that
involves the inflaton (see equation (\ref{eq:Lag_g_K_i})), where the value of
the phase $\vartheta$ can change the rate of the decay of the complex inflaton
into neutrinos. Both mass  and $\nchi$ interaction terms, as expressed by 
Eqs.~(\ref{eq:Lag_mk_equiv}) and (\ref{eq:Lag_IQ_equiv}) remain unchanged 
[see Eqs.~(\ref{eq:Lag_mk}) and (\ref{eq:Lag_IQ})].
On the other hand, this phase could impact the evolution of the
homogeneous background universe, since it appears as part of the total DE
density, as it is shown in equation (\ref{eq:energy_density_X_vartheta}).
However, as it can be seen from  the first slow-roll condition, which in the 
polar base, where we define
\begin{equation}\label{eq:polar_base_Q_ppal}
\Q = \dfrac{(\langle \Q\rangle + \nchi)}{\sqrt 2}e^{i\vartheta/\langle \Q\rangle},
\end{equation}
takes the form [see equation
(\ref{eq:first_slow_roll_condition})],
\begin{equation}\label{eq:first_slow_roll_condition_ppal}
\dfrac{1}{2}\dot\nchi^2 + 
\dfrac{1}{2}\left( 1 + \dfrac{\nchi}{\langle \Q\rangle} 
\right)^2\dot\vartheta^2  \ll \dfrac{1}{2}m^2(\langle Q\rangle + \nchi)^2,
\end{equation}
the phase does not contribute effectively to the DE density, but controls it
indirectly, because the fulfillment of the condition depends on the initial
values of the phase and its velocity. Condition
(\ref{eq:first_slow_roll_condition_ppal}) is fulfilled during the DE dominated
age for most of the initial values of the phase and its velocity, as can be
checked by the evolution of the dynamic system (\ref{eq:dynamic_system_background}).

In particular, for the simplest $\vartheta_{\mathsmaller{ini}} =
\dot\vartheta_{\mathsmaller{ini}} = 0$,  the phase $\vartheta$ remains null
during all the history of the universe, therefore, for these values, the
condition  (\ref{eq:first_slow_roll_condition}) is simplified to the expected
one for the usual case of a real scalar field.

Since in the rest of the present work we will only focus on
(\ref{eq:Lag_IQ_equiv}), the value of the phase will not play a crucial role,
then we can choose the simplest initial condition without losing generality. 
Our model, nonetheless, is completely compatible with different values, as shown
in Appendices A and B, and
although it is not developed here, we believe that a deeper analysis of the
initial conditions could be related to the studies on the problem of
coincidence, as well as to effects beyond the homogeneous limit.

\section{Quintessence quanta and SM particles production}

Because  $\nchi$-particles have the same mass associated with $\Q$ 
they are ultra-relativistic, and thus, right handed pair annihilation 
constitutes a source that can inject an extra degree of freedom 
during the radiation dominated age. Hence, it is necessary to check whether the 
presence of such radiation is compatible 
or not with the predictions of Big Bang Nucleosynthesis (BBN).

In order to do that, we consider standard BBN (SBBN) 
\cite{Wagoner:1966pv,Walker:1991ap,Copi:1994ev} (for a recent review see 
\cite{Cyburt:2015mya}), in which all of the input parameters, namely, the 
number of relativistic degrees of freedom in equilibrium ($g_{*}$), the neutron 
lifetime, the cross-sections of the involved nuclear processes, the mass 
difference between neutrons and protons and the strength of both the weak force 
and gravity, are in accordance with the Standard Model of Particle Physics and 
Einstein gravity.
In SBBN all of those parameters are well determined.  The unique input free 
parameter is the baryon to photon ratio, which determines the primordial 
abundances of the four light nuclei, namely $^4$He, $^3$He, H or D and $^7$Li.
None of them is modified directly in our model,  apart, perhaps,  from 
$g_{*}$. 

Since SBBN assumes a Friedmann-Lema\^itre-Robertson-Walker (FLRW) universe and 
it occurs during the radiation domination age, any increment on $g_{*}$ 
increases the value of the Hubble parameter, $\mathsf H$, consequently, the value of 
the freeze-out temperature of the neutron-to-proton ratio also increases, which 
in turn implies an increment on the final primordial helium abundance.
The same is accomplished if there is some net increase in the total radiation 
energy density due to any process beyond thermal equilibrium.  That is just the 
kind of process of neutrino pair annihilation.

Once the system formed by the neutrinos and $\nchi$-particles goes out of 
equilibrium, the energy density of the latter becomes relevant, otherwise, the 
pair annihilation can be reversed yielding to a net increment of zero in the 
total radiation energy density.
Therefore, to evaluate the total impact on the Hubble parameter, it is 
necessary to determine the out-of-equilibrium radiation production along with 
the one in equilibrium, by evolving the Boltzmann equation for the 
\emph{radiation number density}, $n_\nchi$, as a function of the temperature in 
an FLRW Universe. As the whole process is controlled solely by the 
coupling  $a_c$, and thus by the scale of right handed neutrino masses, the 
analysis of such a process should constrain this parameter in order to avoid 
perturbing the predictions of SBBN through an excess of injected $\nchi$. 
Nevertheless, as we will show hereafter, the process is already so inefficient, 
that no additional constrains are needed on $a_c$, 
in such a way that our model appears as consistent 
with SBBN. Let us proceed next with the detailed analysis.

In order to write the Boltzmann equation, we have to  explicitly calculate the 
collision term, which in turn involves the thermally averaged cross-section 
for the pair annihilation. (For the last calculation we follow 
\cite{Gondolo:1990dk, Srednicki:1988ce}).
We start by calculating the total cross-section for the part of the Lagrangian
(\ref{eq:Lag_IQ_equiv}) that corresponds to only one of the neutrinos, namely
$$
-\Lag_{I \nchi_i} = \dfrac{a_c}{2\sqrt 2}\nchi \bar\nu_i\nu_i.
$$
For this Lagrangian, the total annihilation cross-section  of neutrino pairs 
going to a pair of $\nchi$-particles, calculated in the center of mass frame
(CM), is
\begin{equation}\label{eq:total_cross_section}
  \sigma_{\mathsmaller \nchi} \equiv \sigma_{\bar\nu_i\nu_i \rightarrow \nchi\nchi} = 
  \dfrac{1}{2048\pi s}\dfrac{a_c^4}{v_r(s)\sqrt{\lambda(s,m_k^2)}}F(s),
\end{equation}
where $v_r(s)$ is the relative velocity between the neutrinos and 
\begin{multline}\label{eq:func_F_s}
F(s) = \left[ s + 16m_k^2\left(1 - \dfrac{2m_k^2}{s}\right)  \right]
\log\left[ \dfrac{s + \sqrt{\lambda(s,m_k^2)}}{s - \sqrt{\lambda(s,m_k^2)}}
\right]\\ 
- 2\left( 1 + \dfrac{8m_k^2}{s} \right)\sqrt{\lambda(s,m_k^2)},
\end{multline}
where we have neglected the ultra-relativistic mass $m$ respect to the
non-relativistic neutrino mass $m_k$.
In the previous equations $s$ is a Mandelstam variable, which in the CM
corresponds to $s = 4E^2$, with $E$ the energy of each incoming neutrino and
$\lambda(s,m_k^2)$ is the Mandelstam triangular function, which is given by
$$
\lambda(s,m_k^2) = s(s - 4m_k^2).
$$
Next,  the thermally averaged cross-section becomes
\begin{equation}\label{eq:TACS}
\langle \sigma_{\mathsmaller \nchi} v_r \rangle = \dfrac{a_c^4}{4096\,\pi m_k T\,K_2^2(m_k/T)}\, 
\mathcal I_{\mathsmaller\nchi}(m_k;T),
\end{equation}
where $K_2$ is the modified Bessel function of the second kind of order
$2$, and where we have defined the integral
\begin{equation}\label{eq:integral_I}
\mathcal I_{\mathsmaller\nchi}(m_k;T) \equiv \int_0^1 dx\, \frac{g_{\mathsmaller\nchi}(x)}{x\sqrt x} K_1
\left(\dfrac{2 m_k}{T \sqrt x}\right),
\end{equation}
with $K_1$ the modified Bessel function of the second kind of order $1$,
and $g_{\mathsmaller\nchi}(x)$ the function  coming from (\ref{eq:func_F_s}) after the change of
integration variable
\begin{equation}\label{eq:from_s_to_x}
s \rightarrow  4m_k^2/x,\qquad F(s) \rightarrow 4m_k^2 g_{\mathsmaller\nchi}(x).
\end{equation}

On the other hand,  the out-of-equilibrium  number density for 
the $\nchi$-particles, $n_\nchi$, by means of the Boltzmann equation in an FLRW
Universe,  is given as
\begin{equation}\label{eq:Boltzmann_def}
 \dfrac{1}{a^3} \dfrac{d}{dt}(a^3n_\nchi) = 
 \langle \sigma_{\mathsmaller\nchi} v_r \rangle \left[ n_\nu^2 -  (n_\nu)_{eq}^2\frac{n_\nchi^2}{(n_\nchi)_{eq}^2}\right],
\end{equation}
where $a=a(t)$ is the universal scale factor and the source $n_\nu$ is the 
neutrino number density, which in turn, must be calculated through its own 
Boltzmann equation.

In order to write this last we have to calculate the corresponding collision 
term by considering all the involved processes, namely, annihilation and decay 
of neutrinos into SM particles, together with those of the $\nchi$-channel.
For this we consider the most general Yukawa couplings of our heavy neutrinos, 
$(y^\nu)^{ni}\bar{L}_{n}\widetilde{H}\nu_{i\mathsmaller{R}} + h.c.$, for $i=1,2$ where $L_n$ 
stands for the three standard left handed lepton doublets, with $n=1,2,3$,  and 
$H$ for the Higgs doublet, whose components are denoted as 
$$
H = \doublet{h^+\\ h^0},~~\text{with}~~ 
\widetilde H =i\sigma_2H^\dagger, ~~\text{and}~~ 
L_n = \doublet{\mathcal \nu_{nL}\\ \mathcal \ell_{nL}}~. 
$$
Thus, there are actually two channels for the decay of heavy neutrinos into SM 
particles, 
$$\nu_{i\mathsmaller{R}}\longrightarrow \nu_{nL}h^0\quad\text{ and}\quad\nu_{i\mathsmaller{R}}\longrightarrow 
\ell_{nL}h^+~.$$
On the other hand, the annihilation processes can be written as
$$\nu_i \nu_j \longrightarrow h^{0\dagger}h^0\qquad \text{and} 
\qquad \nu_i\nu_j \longrightarrow h^+ h^-. $$

By considering all of the Yukawa couplings to be about the same order, 
namely $(y^\nu)^{ni}\sim y^\nu$, and because there are six similar processes of 
disintegration of a heavy neutrino, the total decay width is
\begin{equation}\label{eq:total_decay_width}
\Gamma_d = \frac{3}{32\pi}(y^\nu)^2m_k.
\end{equation}

Next, for the annihilation process 
$\bar\nu_i\nu_j \longrightarrow h^{0\dagger}h^0$, the total cross section 
in the (CM) is given as
$$
\sigma_{\mathsmaller H}\equiv
\sigma_{\nu_i\nu_j\rightarrow h^{0\dagger}h^0} = 
\frac{3}{32\pi}\frac{1}{sv_r(s)}\frac{(y^\nu)^4}{\sqrt{\lambda(s,m_k^2)}}T(s),
$$
where
$$
T(s) = \frac{m_k^2 - 2 s} {m_k^2} \sqrt{\lambda(s,m_k^2)} + 3m_k^2\log{\frac{t_{-}}{t_{+}}}, 
$$
with
$$
\frac{t_-}{t_+} = \frac{(s - 2m_k^2) + \sqrt{\lambda(s,m_k^2)}} {(s - 2m_k^2) - \sqrt{\lambda(s,m_k^2)}}.
$$

The thermally averaged cross section involving the three contributions to 
the same channel is then given by
\begin{equation}\label{eq:TACS_H}
\langle \sigma_\mathsmaller{H} v_r\rangle  = 
\frac{9}{4\pi}\frac{(y^\nu)^4}{ m_k T K_2^2(m_k/T)}\mathcal I_\mathsmaller{H}(m_k;T), 
\end{equation}
where as previously, we have defined the integral
$$
\mathcal I_\mathsmaller{H}(m_k;T) = \int_0^1 dx \frac{g_\mathsmaller{H} (x)}{x\sqrt x}K_1\left(\frac{2m_k}{T\sqrt x}\right), 
$$
where
$$
g_\mathsmaller{H}(x) = \frac{2-x}{x}\sqrt{1-x} + \frac{3}{4}\log{\frac{1-\frac{x}{2} + \sqrt{1-x}} {1-\frac{x}{2} - \sqrt{1-x}}},
$$
is the function coming from $T(s)$ after changing the variable as it was defined in (\ref{eq:from_s_to_x}).

It turns out that for the other coannihilation process, 
$\nu_i\nu_j \longrightarrow h^+h^-$, it is fulfilled that
\begin{equation}\label{eq:two_equal_cross_sections}
 \sigma_{\nu_i\nu_j\rightarrow h^+h^{-}} = 
 \sigma_{\nu_i\nu_j\rightarrow h^0h^{0\dagger}} = 
\sigma_{\mathsmaller H},
\end{equation}
due to the $SU(2)$ standard symmetry.
With these, the Boltzmann equation involving both, the Quintessence and Higgs 
channel, becomes
\begin{equation}\label{eq:Boltzmann_equation_n_nu_on_t}
  \frac{1}{a^3}\frac{d}{dt}(a^3n_\nu) = C[T],
\end{equation}
where the collision term is given by
$$
C[T] = \frac{1}{4}\left[\langle\sigma_{\mathsmaller{TOT}} v_r\rangle \left( (n_\nu)^2_{eq} - n_\nu^2 \right) - \Gamma_d n_\nu \right], 
$$
with the total thermally averaged cross section given in terms of the equations 
(\ref{eq:TACS}) and (\ref{eq:TACS_H}) as $$
\langle\sigma_{\mathsmaller{TOT}} v_r\rangle = 
2\langle \sigma_{\mathsmaller\nchi}v_r\rangle + 2\langle \sigma_{\mathsmaller H} 
v_r\rangle, 
$$
wherein the factor of 2 in the $\nchi$-term accounts for the two involved 
Majorana neutrinos [see equation (\ref{eq:Lag_IQ_equiv})], and the second 
factor of 2 is there because of equation (\ref{eq:two_equal_cross_sections}). 
The last term to be defined in the collision term is the neutrino number density 
in equilibrium which is given by
\begin{equation}\label{eq:n_nu_eq}
(n_\nu)_{eq} = 4\pi m_k^2 T K_2(m_k/T).
\end{equation}

After changing time evolution in favor of the temperature in equation 
(\ref{eq:Boltzmann_equation_n_nu_on_t}), which is possible to do during the 
radiation dominated age, one gets
\begin{equation}\label{eq:Boltzmann_equation_n_nu_on_T}
\frac{d}{dT}(a^3n_\nu) = -\frac{M_{pl}}{\pi}\left( \frac{90}{g_{*}(T)} \right)^{1/2}\frac{a^3}{T}C[T],
\end{equation}
where $g_{*}(T)$ is the number of relativistic degrees of freedom in energy
density in equilibrium.

It turns out that $y^\nu$ is always greater than the coupling $a_c$, as can 
be checked by considering that $\langle\Q\rangle \sim m_{\text{pl}}$ and the 
Higgs vacuum expectation value $\langle H\rangle = 246~$GeV into the 
seesaw formula, according to which 
light neutrino mass is given as $m_\nu \sim m_{\mathsmaller D}^2/m_k$. For an 
$m_\nu$  well within the observed bounds \cite{Tanabashi:2018oca}, which for 
the 
heaviest of the light states are given as
 $ 5\times 10^{-2}~\text{eV}  ~\lesssim~  m_\nu  ~\lesssim~ 
10^{-1}~\text{eV}~$,
we arrive to
\begin{equation}\label{eq:bounds_to_ratio_y_to_ac}
\frac{(y^\nu)^2}{a_c}\sim (1.4 - 2.8 )\times 10^{4}~,
\end{equation}
with the largest values corresponding to the cosmological neutrino mass bound, 
and the lowest to atmospheric neutrino oscillation scale.  
This means that both, the decomposition 
channel and the co-annihilation channel, dominate over that of $\nchi$-quanta 
production.
By taking for instance,  $y^\nu \sim 1$, we then get $a_c \sim 10^{-4}$ in 
accordance with our assumptions.

\begin{figure}[h]
\begin{center}
\includegraphics[width=\columnwidth]{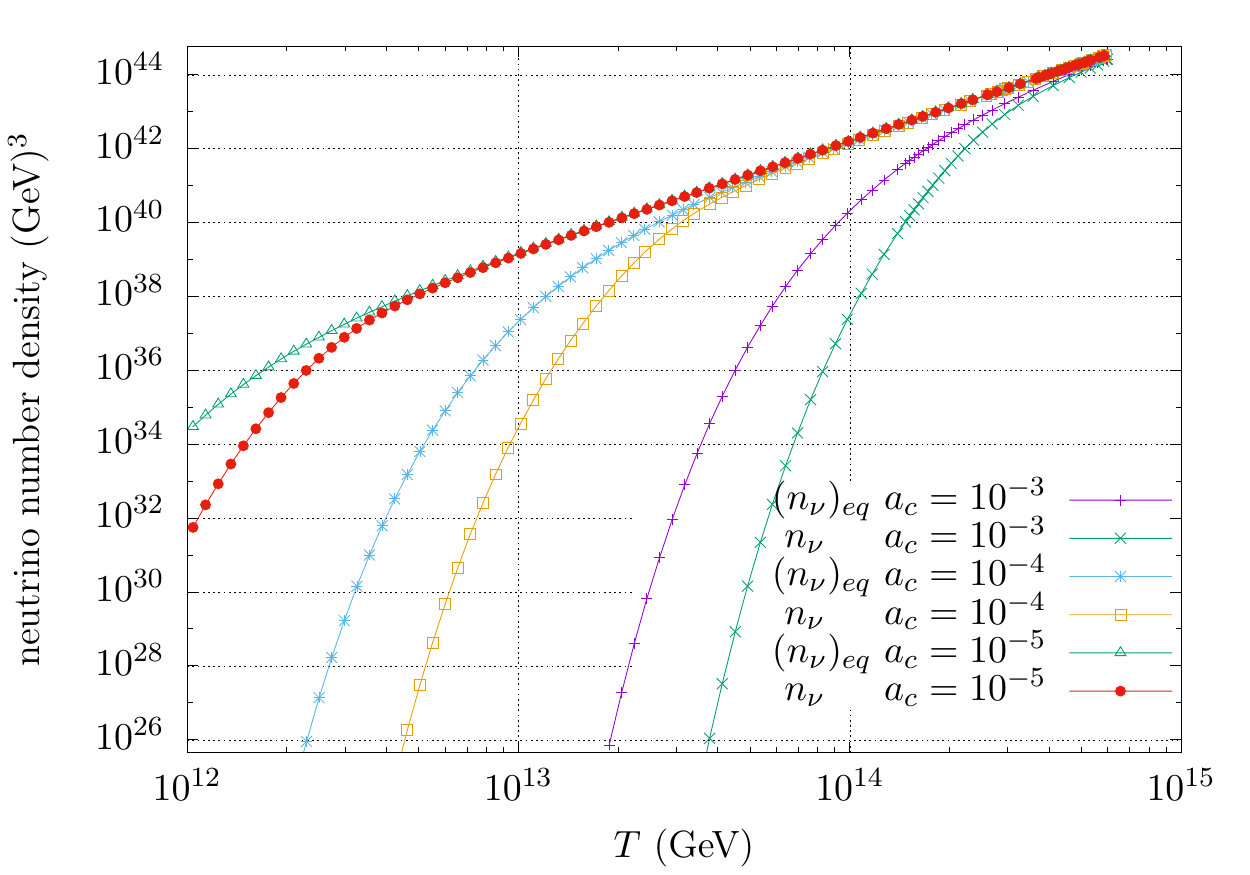}
\caption{The neutrino number density $n_{\nu}$ varying on temperature driven by 
equation (\ref{eq:Boltzmann_equation_n_nu_on_T}) and the equilibrium number 
density given in (\ref{eq:n_nu_eq}). Notice that the system goes 
out-of-equilibrium at early times but the number density gets suppressed 
strongly  due to the decay term proportional to (\ref{eq:total_decay_width}). }
\label{fig:Boltzmann_eq_n_nu_evolution}  
\end{center}
\end{figure}

Next, in order to calculate the evolution of the number density $n_\nu$, we sall set $y^\nu = 1$ from now on.
Thus, by numerical evolving equation 
(\ref{eq:Boltzmann_equation_n_nu_on_T}), we found that the number density 
$n_\nu$ never override that of equilibrium $(n_\nu)_{eq}$, as it is shown in 
figure \ref{fig:Boltzmann_eq_n_nu_evolution},  in accordance to which, the 
system evolves in thermal equilibrium at early times and then leaves equilibrium 
to get highly suppressed due to the decay channel characterized by 
(\ref{eq:total_decay_width}).  

\begin{figure}[h]
\begin{center}
\includegraphics[width=\columnwidth]{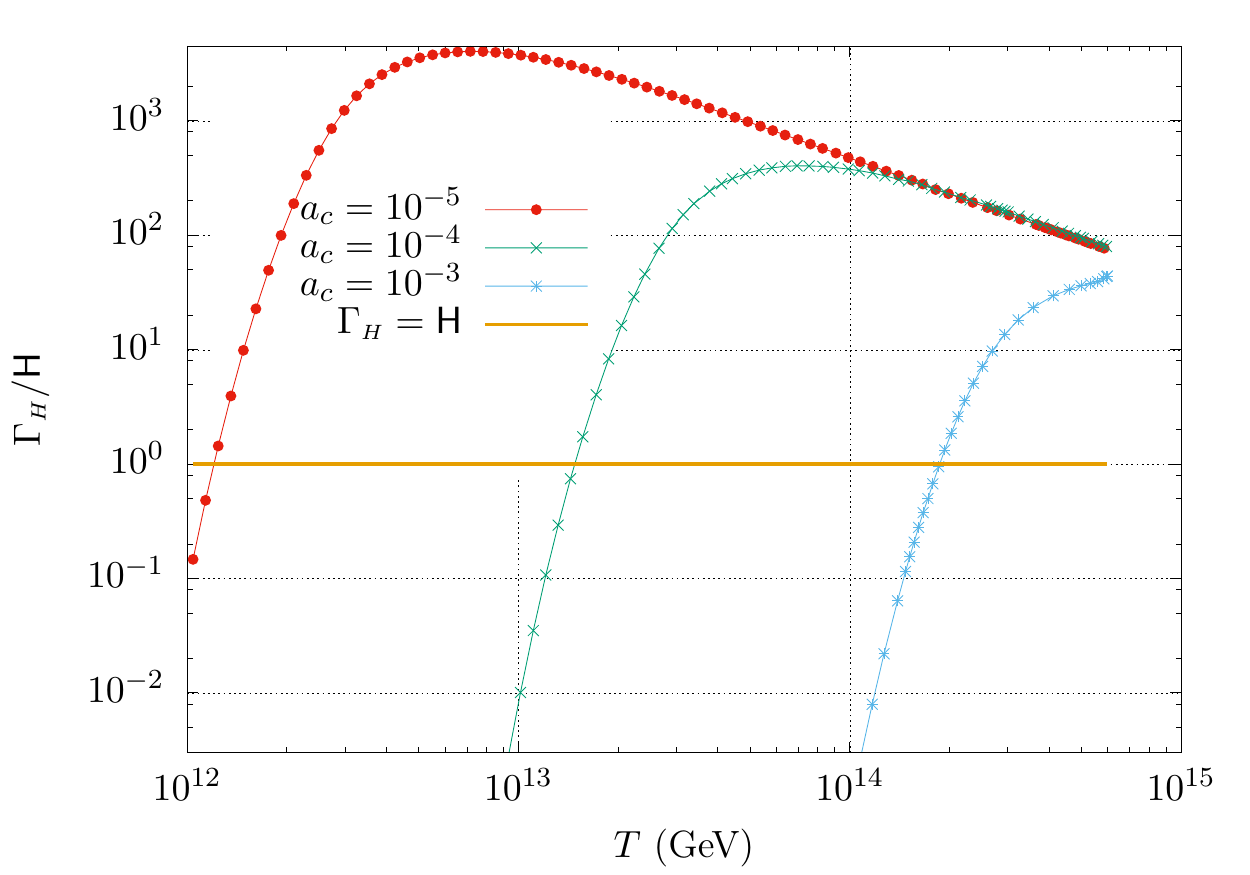}
\caption{The out-of-equilibrium condition given in
Eq.~(\ref{eq:out_of_equilibrium_condition_Higgs}) for some values of the 
parameter $a_c$.  As stated in the text,  the greater the mass $m_k$ the earlier 
the beginning of the out-of-equilibrium epoch, due to the decay processes into 
SM particles.
}
\label{fig:Equilibrium_condition_Higgs}  
\end{center}
\end{figure}

In figure \ref{fig:Equilibrium_condition_Higgs} we plot, for a few values of the 
coupling $a_c$, the out-of-equilibrium condition 
\begin{equation}\label{eq:out_of_equilibrium_condition_Higgs}
\Gamma_{\mathsmaller H} \equiv 2\times \langle \sigma_{\mathsmaller H} v_r \rangle\times n_\nu \lesssim \mathsf H,
\end{equation}
where $\Gamma_{\mathsmaller H}$ 
is the neutrino interaction rate for the Higgs channel and $\mathsf H$ is the 
Hubble parameter. As it is shown there, because of the decay of neutrinos into 
Higgs and leptons, the greater the coupling $a_c$ (and so the mass $m_k$), the 
earlier the out-of-equilibrium epoch. This also shows that, as expected, 
neutrino into SM fields coannihilation is efficient enough at higher 
temperatures as to thermalize the heavy neutrinos.

\begin{figure}[h]
\begin{center}
\includegraphics[width=\columnwidth]{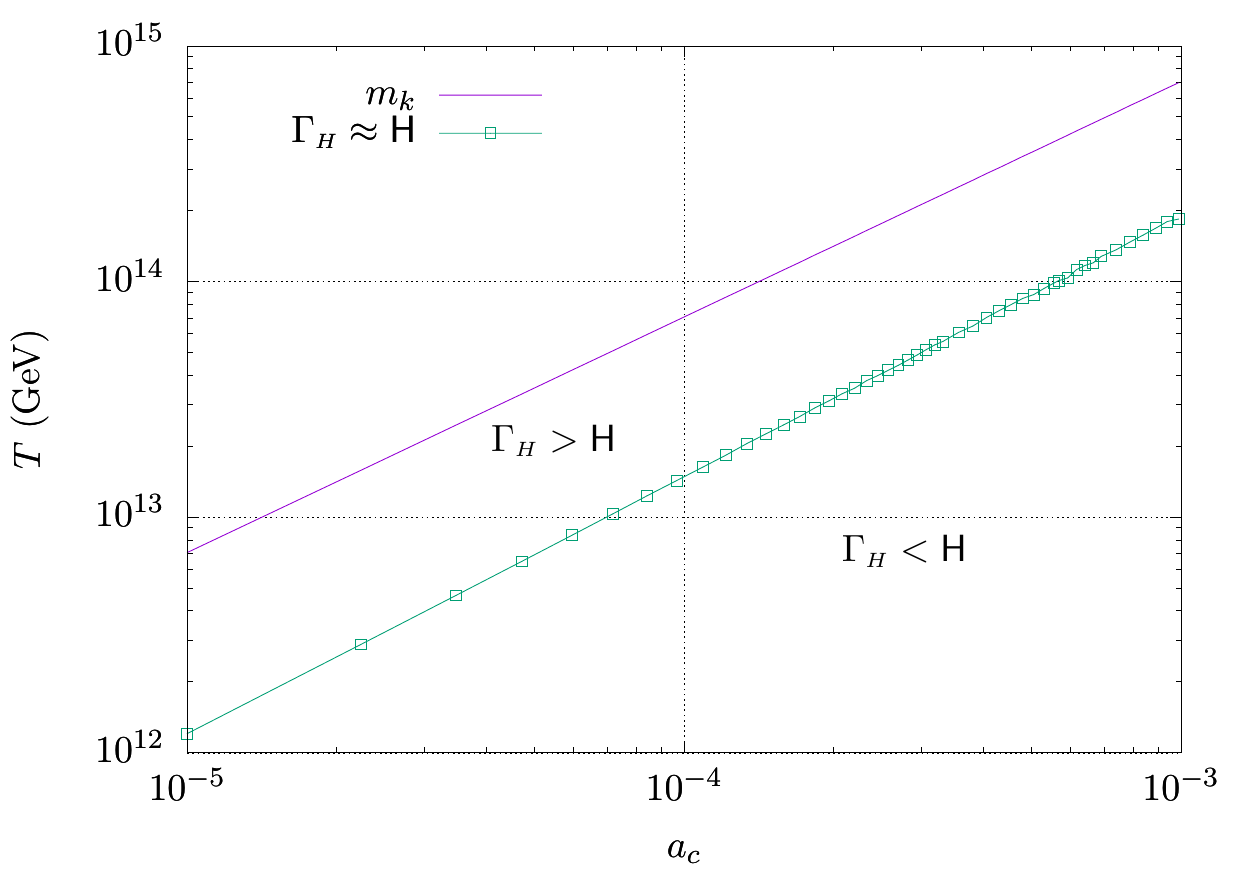}
\caption{
The behaviour of the population of Majorana sterile neutrinos as a function of 
the parameter $a_c$ at a given temperature. As stated in the text, between $T_r 
\sim 10^{15}$ GeV and $T \approx m_k$, the number density of neutrinos 
corresponds to that of radiation in thermal equilibrium, between $T\approx m_k$ 
and the temperature of equality $\Gamma_\mathsmaller{H}\approx \mathsf H$, the 
system becomes non-relativistic but still keeps in thermal equilibrium. Below 
this line the system goes out-of-equilibrium and the population of neutrinos 
gets suppressed due to the processes of annihilation and decay populing the 
universe with SM particles.     }
\label{fig:Tout_vs_ac}
\end{center}
\end{figure}

In figure \ref{fig:Tout_vs_ac} we illustrate the behavior  of the system in the 
space of the temperature versus the parameter $a_c$. As said before,  
inflaton decays into neutrinos $\nu$ and reheats the Universe at temperature 
$T_r \sim 10^{15}$ GeV, below this temperature  and above the upper line that 
stands for the value of the mass $m_k$,  the population of neutrinos behave like 
pure radiation in thermal equilibrium and stays that way until the temperature 
drops into the region below the line $m_k$ and above the line 
$\Gamma_\mathsmaller{H} \approx \mathsf H$, in which the system becomes 
non-relativistic but still keeps in thermal equilibrium. Below the bottom line 
the system goes out-of-equilibrium, and the population of neutrinos decreases 
due to the co-annihilation into Quintessence and Higgs pairs, as well as the 
decay into Higgs and leptons. As a final notice, we mention that adding extra 
Higgses to the model will add to the number of coannihilation channels, 
introducing an overall factor to the corresponding rates which will not affect 
our conclusions above.

Turning back to the Boltzmann equation (\ref{eq:Boltzmann_def}), notice that, 
because we are interested in maximizing the production of $\nchi$-quanta, which 
states the worst possible scenario for the model, and because $(n_\nu)_{eq} 
\gtrsim n_\nu$ [see fig. \ref{fig:Boltzmann_eq_n_nu_evolution}],  we can choose 
$(n_\nu)_{eq}$ instead of $n_\nu$ in the collision term,  and we can neglect the 
ratio $n_\nchi^2 / (n_\nchi)_{eq}^2$ which accounts for a tiny fraction of 
$n_\nchi$ produced due to the inverse proccess 
$\nchi\nchi\rightarrow\bar\nu\nu$, in this way we overestimate the production of 
$\nchi$-quanta. As we will shown next, this approximation will be enough to 
stablish the cosmological consistency of the model, because the result does not 
conflict with the requirements of Big Bang Nucleosynthesis.
Then the equation (\ref{eq:Boltzmann_def}) becomes
\begin{equation}\label{eq:Boltzmann_overestimated}
 \dfrac{1}{a^3} \dfrac{d}{dt}(a^3n_\nchi) = 
 \langle \sigma_{\mathsmaller\nchi} v_r \rangle (n_\nu)_{eq}^2.
\end{equation}
 By using equations (\ref{eq:TACS}) and (\ref{eq:n_nu_eq}), 
the Boltzmann equation (\ref{eq:Boltzmann_overestimated}) becomes
\begin{equation}\label{eq:Boltzmann_form_t}
 \dfrac{1}{a^3} \dfrac{d}{dt}(a^3n_\nchi) = \dfrac{\pi}{256}m_k^3 a_c^4 T\, 
\mathcal I_\mathsmaller{\nchi}(m_k;T).
\end{equation}
As before, after changing time evolution in favor of the temperature,  the Boltzmann equation 
(\ref{eq:Boltzmann_form_t})
becomes 
\begin{equation}\label{eq:Boltzmann_form_T}
\dfrac{d}{dT}(a^3 n_\nchi) = - \dfrac{M_{pl}}{256}\left( \dfrac{90}{g_{*}(T)}  
\right)^{1/2} m_k^3 a_c^4 \dfrac{a^3}{T^2} \mathcal I_{\mathsmaller\nchi}(m_k;T).
\end{equation}

Since the Universe is cooling, we perform the integration at both sides backward
in $T$, from $T_{out}$ to a certain temperature $T' < T_{out}$, so, we  have
\begin{multline}\label{eq:integration_01}
\int_{(a^3 n_\nchi)(T_{out})}^{(a^3 n_\nchi)(T')} d(a^3 n_\nchi) =\\ 
-\dfrac{M_{pl}}{256}\sqrt{90}m_k^3 a_c^4\,\int_{T_{out}}^{T'} dT 
\dfrac{a^3(T)}{\sqrt{g_{*}(T)}T^2}\mathcal I_{\mathsmaller\nchi}(m_k;T),
\end{multline}
where in the RHS, we have written explicitly the universal scale factor
dependence on $T$, such a dependence, during the radiation dominated age,  is
given by
\begin{equation}\label{eq:scale_factor_on_T}
a(T) = \dfrac{b_0}{g_{*s}^{1/3}(T)T},
\end{equation}
where $b_0$ is a constant and $g_{*s}(T)$ is the number of relativistic degrees
of freedom in entropy density in equilibrium.

\begin{figure}[h]
\begin{center}
\includegraphics[width=\columnwidth]{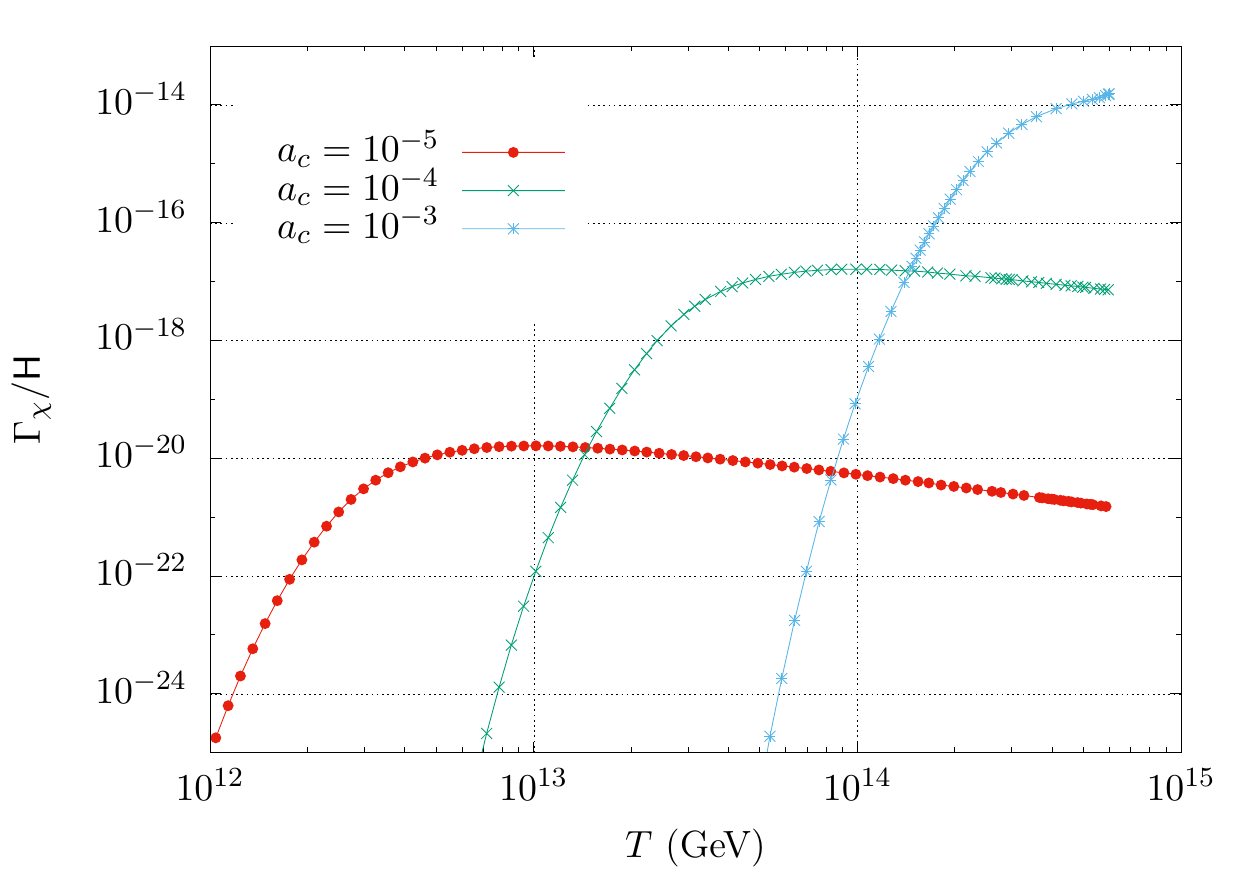}
\caption{The out-of-equilibrium condition given in
Eq.~(\ref{eq:equilibrium_condition_Chi}) for some values of the parameter $a_c$.  
As stated in the text,  the integral (\ref{eq:integral_I}) is very suppressed, 
hence the system $\bar\nu\nu\leftrightarrow\chi\chi$ is always out
of equilibrium, even for temperatures as high as of the one for reheating. 
}
\label{fig:Gamma_chi_over_H}
\end{center}
\end{figure}

When the cooling Universe reaches the temperature $T_{out}$ the density 
$n_\nchi$
starts to increase, i.e. the system $\bar\nu\nu\leftrightarrow\nchi\nchi$ goes out of equilibrium, which is true
whenever
\begin{equation}\label{eq:equilibrium_condition_Chi}
  \Gamma_{\mathsmaller\nchi} \equiv 
  2\times\langle \sigma_{\mathsmaller\nchi} v_r \rangle \times n_\nu
  ~\lesssim~ 
\mathsf H,
\end{equation}
where $\Gamma_{\mathsmaller\nchi}$ is the neutrino interaction rate of the 
channel, which can be calculated by using (\ref{eq:TACS}) and the numerical 
output of (\ref{eq:Boltzmann_equation_n_nu_on_T}).
It turns out that for any value of $T \lesssim T_r$, the integral
(\ref{eq:integral_I}) is very suppressed and so is the rate $\Gamma_{\mathsmaller\nchi}$ as it is
shown in the figure \ref{fig:Gamma_chi_over_H}. Then the inequality
(\ref{eq:equilibrium_condition_Chi}) is always fulfilled and we can use the
temperature $T_{out} \sim T_r $ as the lower limit to obtain a good estimate of 
the integral that appears in the RHS of Eq.~(\ref{eq:integration_01}).

Furthermore, as the initial state of the $\nchi$-field is one of pure vacuum,
and this is not coupled to the inflaton,
there are not initial quanta, consequently, we can impose the
condition
$$
(a^3n_\nchi)(T_{out}) = 0, 
$$
which jointly to Eq.~(\ref{eq:scale_factor_on_T}) allows expressing  the 
integral in Eq.~(\ref{eq:integration_01}) as
\begin{equation}\label{eq:integration_02}
n_\nchi(T') = N a_c^7  g_{*s}(T') T^{'3}\int_{T'}^{T_r} 
\dfrac{dT}{T^5}\dfrac{\mathcal I_{\mathsmaller\nchi}(m_k;T)}{g_{*s}(T)\sqrt{g_{*}(T)}} ,
\end{equation}
where $N$ is a constant factor given by
\begin{equation}\label{eq:N_factor}
N = 2\times  \dfrac{M_{pl}}{512}\sqrt{45} \langle\Q\rangle^3,
\end{equation}
and where we have multiplied it by $2$ because there are two Majorana neutrinos
involved [see equation (\ref{eq:Lag_IQ_equiv})].  

By considering $g_{*s} \sim g_{*n}$, whit $g_{*n}$ the relativistic degrees of
freedom in number density in equilibrium,  the integral
(\ref{eq:integration_02}) can be written as
\begin{equation}\label{eq:integration_03}
n_\nchi(T') = n_r(T') \times f(T'),
\end{equation}
where $n_r(T')$ is the relativistic number density in equilibrium, given by
\begin{equation}\label{eq:rela_number_densi}
n_r(T') = \dfrac{\zeta(3)}{\pi^2}g_{*n}(T')T'^3, 
\end{equation}
with $\zeta(3)$ the Ap\'ery's constant, and where
\begin{equation}\label{eq:function_f}
f(T')=   N a_c^7  \dfrac{\pi^2}{\zeta(3)}\int_{T'}^{T_r} 
\dfrac{dT}{T^5}\dfrac{\mathcal I_{\mathsmaller\nchi}(m_k;T)}{g_{*s}(T)\sqrt{g_{*}(T)}}.
\end{equation}

\begin{figure}[h]
\begin{center}
\includegraphics[width=\columnwidth]{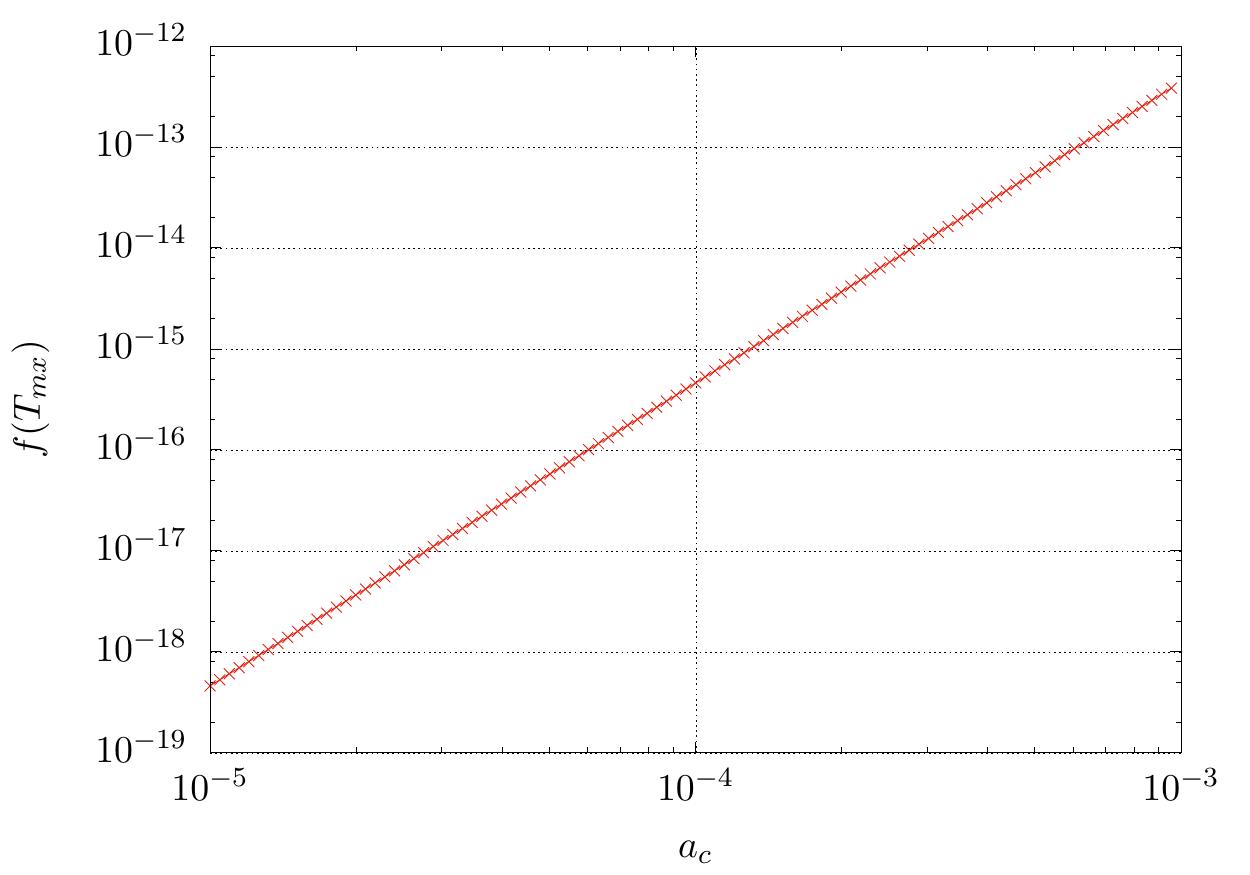}
\caption{The maximums of the function $f(T)$ given in (\ref{eq:function_f}) for
different values of the parameter $a_c$. Notice that the integral is always 
less than the unit and so the increase in $n_{\mathsmaller{\text{TOT}}}$ given 
in Eq.~(\ref{eq:n_tot}) is negligible.} \label{fig:the_function_f}
\end{center}
\end{figure}

By means of equation (\ref{eq:integration_03}) we write the total
relativistic number density of our model $n_{\mathsmaller{\text{TOT}}}$, in
terms of (\ref{eq:function_f}) as
\begin{equation}\label{eq:n_tot}
n_{\mathsmaller{\text{TOT}}}(T') = n_r(T')(1 + f(T')).
\end{equation}
The integral (\ref{eq:function_f}) can be calculated numerically for different
values of the $a_c$ parameter, with the result that for each value of the
latter, the integral depends smoothly on the temperature and it is easy to
maximize.

Since $n_r(T)$  is a growing monotonic function, it is enough to know whether,
for certain $a_c$, the value of $f(T_{mx})$ exceeds that of $n_r(T_{mx})$, where
$T_{mx}$ is the temperature that maximizes the integral (\ref{eq:function_f}).
What we found is that $f(T_{mx})$ is always several orders of magnitude below
one for any value of $a_c$ within the range we are interested on, as shown in figure \ref{fig:the_function_f},
so the increase in the total relativistic number density of 
$\nchi$ particles due to the co-annihilation of right handed neutrinos is of no 
cosmological consequences. Clearly, once neutrino decay into SM fields is 
switched on, the actual $\nchi$ would be much smaller that the value  we have  
just calculated.  
The model, to this extent, appears consistent with 
the cosmological constraints.

\section{Thermal corrections to quintessence mass}
As stated earlier, the $\Q$ field would actually couple to light active 
Majorana neutrinos, $\nu_l$, that emerge 
from the seesaw mechanism. By setting the corresponding seesaw mixing into 
Eq.~(\ref{eq:Lag_IQ_equiv}), we get the effective coupling
$-\Lag_{I \nchi} = \lambda\nchi \bar\nu_l\nu_l+ \dots $,
where 
\begin{equation*}
\lambda = \frac{a_c}{2\sqrt{2}}\left( \frac{m_{\mathsmaller D}}{m_k} 
\right)^2~.
\end{equation*}
Then, the thermal contribution to the quintessence potential due to the 
cosmological neutrino background (see for instance \cite{Baldes:2016rqn}) is 
given by 
\begin{equation*}
V^{\mathsmaller T}(\Q,T) = \frac{g}{48}\lambda^2 T^2 \langle\Q\rangle^2~,
\end{equation*}
where $g$ are the degrees of freedom of the Majorana neutrinos coupled to the 
scalar field and $T$ corresponds to the neutrino bath temperature, $T_\nu$,
which scales as
\begin{equation*}
  T_\nu  = T_{\nu,0}\left(\frac{a_0}{a} \right)= (1+z)~T_{\nu,0}~,
\end{equation*}
with $a_0$ the current value of the scale factor $a$,  $z$ the cosmological  
redshift and $T_{\nu,0}$ the current neutrino temperature, which
is related to the current CMB temperature by
$T_{\nu,0} = \left( {4} / {11} \right)^{1/3}T_{\gamma,0}~,$
and so (see \cite{Tanabashi:2018oca})  
\begin{equation*}
T_{\nu,0} = 1.676\times 10^{-4}~\text{eV}.
\end{equation*}

Thus, the effective potential, defined with the addition of above thermal 
correction, becomes 
$$V^{\text{eff}}(\Q,T) = \frac{1}{2} M_{\text{eff}}^2\langle\Q\rangle^2~,$$
where the effective mass is defined as
\begin{equation*}
M_{\text{eff}}^2  = m^2 + m_{\mathsmaller{Th}}^2~,
\end{equation*}
with the \emph{thermal mass} given by
\begin{equation*}
m_{\mathsmaller{Th}}(T_\nu) = \sqrt\frac{g}{24}~\lambda T_\nu~.
\end{equation*}
Clearly, $m_{\mathsmaller{Th}}$ should scale down
as $m_{\mathsmaller{Th}}=(1+z)~m_{\mathsmaller{Th},0}~$, where, obviously, 
its current value $m_{\mathsmaller{Th},0} = 
m_{\mathsmaller{Th}}(T_{\nu,0})$.

By using the corresponding mass expressions, it is 
straightforward 
to write the effective coupling in terms of Yukawa 
couplings and vacuum scalar values, as
$$
\lambda = \frac{1}{2\sqrt 2}\frac{(y^\nu)^2}{a_c}
\left(\frac{ \langle H\rangle }{\langle\Q\rangle } \right)^2~.
$$
Then, by using that 
$\langle H\rangle / \langle\Q\rangle \approx 2.05\times 10^{-17}$,
and the range limits imposed from neutrino masses, given in
Eq.~(\ref{eq:bounds_to_ratio_y_to_ac}), we obtained that the effective coupling 
should be in the narrow range
\begin{equation*}
\lambda \approx (2.05  - 4.1) \times 10^{-30}~.
\end{equation*}
Considering the contribution of the three active neutrinos  to the  
degrees of freedom, we get  $g = 5.25$, and thus, the current thermal mass 
should be
\begin{equation}\label{eq:bounds_to_mT0}
m_{\mathsmaller{Th},0} \approx (1.6 - 3.2) \times 10^{-34}~, 
\end{equation}
which is just in the same range of the needed value to source the cosmological 
constant nowadays  [see Eq.~(\ref{eq:scalar_mass_required})], by 
$\rho_{\mathsmaller{DE}} = \frac{1}{2}M_{\text{eff},0}^2\langle \Q\rangle^2$. As a further note,  since 
$\lambda$ depends linearly on light neutrino mass, $m_{\mathsmaller{Th},0}$ 
would get closer to the required $M_{\text{eff},0}$ for the larger masses. In 
particular, for  $m_\nu\sim 0.18~\text{eV}$,  one already gets 
$m_{\mathsmaller{Th},0}\sim 5.8\times 10^{-34}\text{~eV}$. This would then be 
the largest allowed value for the neutrino mass in our model.

\subsection{Slow-roll condition}
Next, we consider the redshift of the thermal mass to explore 
about the evolution of the slow-roll condition to ensure that the effective 
mass would remain smaller than the Hubble parameter along Universe history, 
which kept $\Q$ field behaving as a true DE. As it should be clear, the main 
concern would be just the $a^{-1}$ scaling of thermal mass value along the 
ages. 

First, during radiation dominated age, the Hubble 
parameter evolves as
\begin{equation*}
\mathsf H^2 = \mathsf H_0^2  \Omega_{\mathsmaller{R},0}  \left( \frac{a}{a_0} \right)^{-4}, 
\end{equation*}
then,  during that age, the quotient between 
the thermal mass and the Hubble parameter becomes
\begin{equation*}
\frac{m_{\mathsmaller{Th}
}}{\mathsf H} = 
\frac{ m_{\mathsmaller{Th},0} }{\mathsf H_0 
\sqrt{\Omega_{\mathsmaller{R},0}}}~\left( \frac{a}{a_0} \right).
\end{equation*}
By using  $\mathsf H_0 = 1.44 \times 10^{-33}$ eV, 
$\Omega_{\mathsmaller{R},0} = 1.4\times 10^{-3}$ together with 
(\ref{eq:bounds_to_mT0}), we arrive to
\begin{equation*}
 \frac{m_{\mathsmaller{Th}}}{\mathsf H}\eval_{\mathsmaller{RAD}}
\approx ( 3 - 6 ) \left( \frac{a}{a_0} \right),\quad a \leq a_{eq}~,
   \end{equation*}
where $a_{eq}$ is the scale factor at the radiation matter equality. 
When $a = a_{eq}$, we have  that 
$\left( {a_{eq}}/{a_0} \right) =  4.45 \times 10^{-3}~.$ With this value,  at 
the end of the radiation age 
\begin{equation*}
\frac{m_{\mathsmaller{Th}}}{\mathsf H}\eval_{a=a_{eq}}
\approx (1.33 - 2.66) \times 10^{-2}~, 
 \end{equation*}   
and thus,  we see that the slow-roll condition does is 
fulfilled during all the radiation age.

During the matter-dominated era, the Hubble para\-meter rather  evolves as
\begin{equation*}
\mathsf H^2 = \mathsf H_0^2  \Omega_{\mathsmaller{M},0}  \left( \frac{a}{a_0} 
\right)^{-3},
\end{equation*}
then, the ratio of interest scales as
\begin{equation*}
\frac{m_{\mathsmaller{Th}
}}{\mathsf H} = \frac{m_{\mathsmaller{Th},0}
 }{\mathsf H_0 
\sqrt{\Omega_{\mathsmaller{M},0}}  } \left( \frac{a}{a_0} \right)^{1/2} ~.
\end{equation*}
By taking  $\Omega_{\mathsmaller{M},0} = 0.3142$, one gets
\begin{equation*}
\frac{m_{\mathsmaller{Th}}}{\mathsf H}\eval_{\mathsmaller{MAT}}
\approx (0.2 - 0.4)\left( \frac{a}{a_0} \right)^{1/2}~. 
\end{equation*}
Considering that in our model [see Eq.~(\ref{eq:Hubble_efectivo})], at the epoch of the transition to DE domination 
$\left({a_{\mathsmaller{DE}}}/{a_0}  \right) \sim  0.45$,
we roughly estimate that 
\begin{equation*}
\frac{m_{\mathsmaller{Th}}}{\mathsf H}\eval_{a = a_\mathsmaller{DE}}\sim 
(0.13-0.27)~.
\end{equation*}
This means that even during the era of the matter domination, the condition of 
slow-roll is still fulfilled, and this would remain so until today 
since thermal mass keeps scaling down as the Universe expands. As a matter of 
fact, Eq.~(\ref{eq:bounds_to_mT0}) implies that nowadays
$$
\frac{m_{\mathsmaller{Th,0}}}{\mathsf H_0} = (0.11-0.22)~.
$$

Last, but not least,  by considering our above mentioned upper bound on the 
neutrino mass ($m_\nu=1.8\times 10^{-1}$~eV), which saturates DE density, such 
that $M_{\text{eff},1.8}\approx m_{\mathsmaller{Th,0}}$, we note that such a case  
provides the natural upper bound on the slow-roll condition.  For this given 
value, we can make a more careful estimate of the evolution of the mass to Hubble ratio
\begin{equation}\label{eq:ratio_Meff_to_H}
R = \frac{M_{\text{eff},1.8}}{\mathsf H_{\text{eff}}}~, 
\end{equation}
by observing that  DE density would then scale as
$$
\rho_{\mathsmaller{DE},1.8} = 
\frac{1}{2}M_{\text{eff},0}^2\left(\frac{a_0}{a}  \right)^2\langle\Q\rangle^2~,
$$
corresponding to a DE relative density parameter  
$$
\Omega_{\mathsmaller{DE},1.8} = 
\frac{\rho_{\mathsmaller{DE},1.8}}{\rho_{\text{crit}}}  
=\Omega_{\mathsmaller{DE},0}\left(\frac{a_0}{a}\right)^2~,
$$
where, by using that  $\rho_{crit} = 3.69\times 10^{-47}~\text{GeV}^4$ it is 
obtained that 
$
\Omega^{\text{eff}}_{\mathsmaller{DE},0} \approx 0.685
$,
as expected under our construction.

\begin{figure}[h]
\begin{center}
\includegraphics[width=\columnwidth]{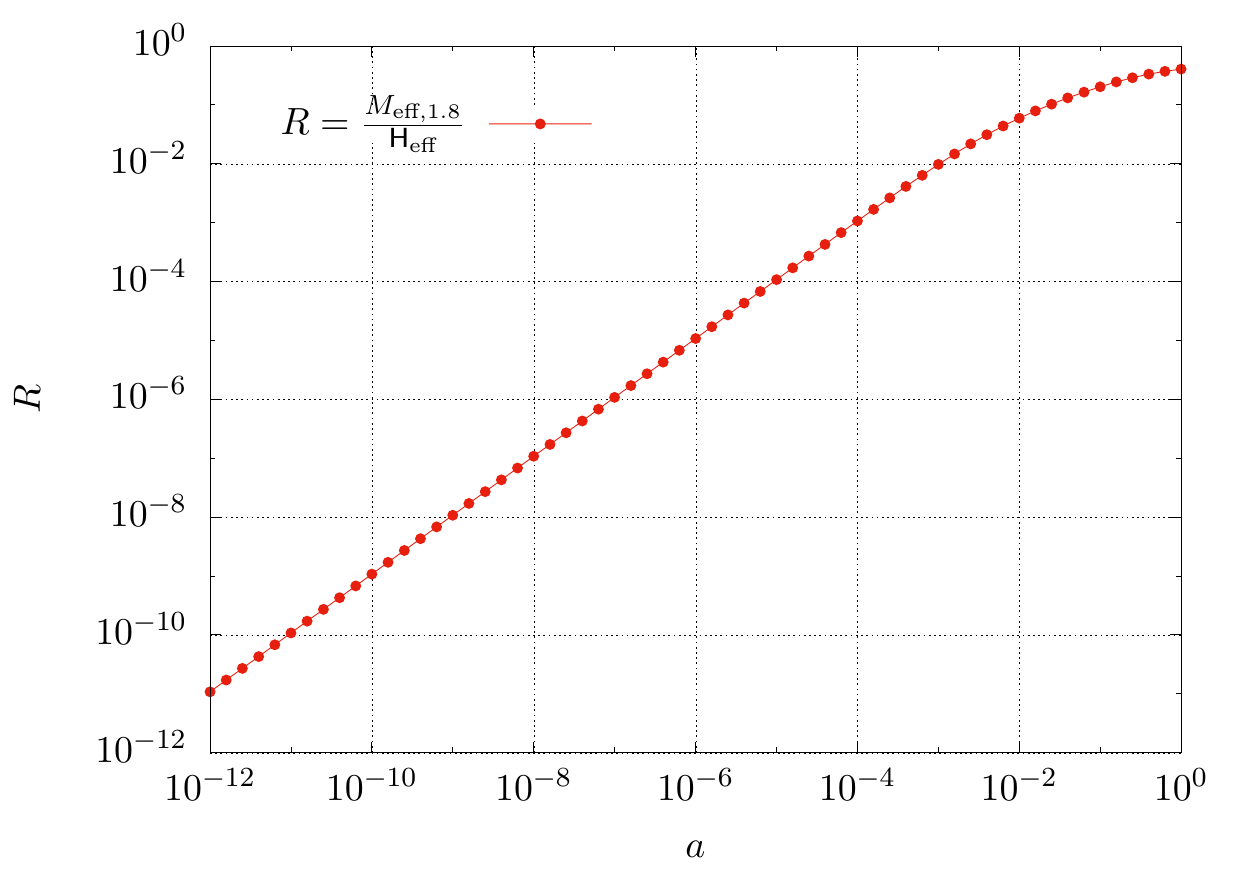}
\caption{The ratio $R$ as defined in Eq.~(\ref{eq:ratio_Meff_to_H}). The 
slow-roll condition is accoplished during the whole life of the Universe. Lower 
values are expected when $m_\nu \rightarrow 0.5\times 10^{-2}$. }
\label{fig:ratio_R}  
\end{center}
\end{figure}

On the other hand, using the actual observed values of 
$\Omega_{\mathsmaller{R},0}$ and $\Omega_{\mathsmaller{M},0}$ mentioned above, 
the effective Hubble parameter to calculate $R$ at any given $a$ can be written 
as
\begin{multline}\label{eq:Hubble_efectivo}
  \mathsf H_{\text{eff}}^2 = \mathsf H_0^2 ~
  \left[ 1.4\times 10^{-3}\left( \frac{a_0}{a} \right)^4 + 0.314 \left( 
\frac{a_0}{a} \right)^3\right.\\
+ \left. 0.685\left( \frac{a_0}{a} 
\right)^2\right]~.
\end{multline}
The numerical evolution of the ratio is plotted in figure \ref{fig:ratio_R}, 
from which we can see that the slow-roll condition is accomplished during the 
whole life of the Universe.

\section{Summary and concluding remarks}

We have presented a cosmological model that unifies early inflation and late 
accelerated expansion, driven by a quintessence field, where both cosmological 
scalar fields belong to the degrees of freedom of the same fundamental field 
representation, $\Phi$, of the $SO(1,1)$ symmetry. This
symmetry, as it is usual in particle physics model building, in particular 
in the construction of the Standard Model, is the guiding principle that 
dictates and governs the dynamics of the system. It is really interesting that 
such a  simple principle allows reproducing chaotic type potentials for both 
inflation and DE, which are derived from considering all possible bilinear 
field operators based on $\Phi$ that are invariant under the symmetry.
As a matter of fact, the field system of the model can be 
rewritten in terms of two scalar fields with and independent evolution, which 
in the cosmological setup will fall down on simple mass type potential. Upon 
fine-tuning, one can easily understand the reason why one of such fields 
breaks down the slow-roll condition at large scale, ending inflation, whereas 
the other stays trapped in a false vacuum configuration that we see as a 
cosmological constant nowadays.

The need for reheating after inflation, which requires the coupling of the 
inflaton to matter fields, is fulfilled by introducing a set of fermions
which, in order to be consistent with the symmetry, belong to a doublet and 
singlet of $SO(1,1)$. 
Enforcing the symmetry to build the Yukawa couplings as also invariant terms 
has two outstanding implications. First,  
since the cosmological field does not belong to the 
Standard Model particle sector,  neither the new fermions will, and thus 
they are naturally identified as right handed neutrinos. Second,  the invariant 
couplings among $\Phi$ and the fermions do provide the  appropriate inflaton 
couplings to allow inflaton decay and reheating, but furthermore, they also 
mean that right handed neutrinos would couple to the cosmological DE field.
Without any further assumption, beyond the use of symmetries, our model 
introduces a way to naturally understand the existence of large sterile 
Majorana neutrino masses as sourced by DE, which, on the other hand, is a need 
for the standard seesaw mechanism to work. The last is the simplest known 
mechanism that provides very small masses to the standard neutrinos, required to 
explain neutrino oscillation phenomena. 

Here, we have studied in some detail the mechanism contained in the $SO(1,1)$ 
cosmological Unification model that is beneath the generation of neutrino 
masses. Our analysis shows that the origin of the mass is independent of the 
field phases and their dynamics. However, it may not be the only possible 
mechanism in nature, as the $SO(1,1)$ symmetry does not prohibit to write an 
independent mass associated to any singlet fermion. Such a mass seems 
unnatural since there is no a priory mass scale associated to it, an issue 
already present in the seesaw. Nevertheless, as we have 
argued, such a mass can easily be removed if additional global symmetries are 
involved in the fermion sector. In such a scenario, DE arises as the natural 
source of such a neutrino mass, through its false vacuum energy that supports 
current accelerated expansion of the Universe. As this last has a large scale, 
then it comes naturally that the right handed neutrinos would have masses in the 
$10^{14}$~GeV, scale, or so.

Our study has also looked upon the possible impact that the model 
and in particular quintessence quanta, $\nchi$ may have in the thermal history 
of the Universe. The inflaton in the model does not couple to quintessence 
field, and thus, it does not inject entropy through that channel upon decay. As 
a matter of fact, the only allowed decay channel for the inflaton is for its 
decay into the heavy right handed neutrinos, which, eventually would create the 
primordial plasma through Higgs and Standard Model lepton couplings, of the 
form $\bar L \widetilde{H} N$. After this we expect Standard thermal history to 
proceed as 
usual, but for the possible contributions to entropy that the right handed 
neutrinos would inject back into the Universe, in the form of quintessence 
quanta through out-of-equilibrium co-annihilation processes, 
$\bar\nu\nu\rightarrow \nchi\nchi$. To further estimate this effect, we have  
calculated the  thermally averaged cross section for the process, which 
depends on the same Yukawa coupling that provides neutrino masses, $a_c$. 
As discussed in the paper, the numerical integration of the  Boltzmann 
equations with $a_c$ varying on a wide range of values shows that the process 
is so suppressed that the total amount of injected quintessence quanta number 
density is negligible. This clearly indicates that the model, without any 
further constraints or assumptions, remains consistent with the conditions 
required for a successful Big Bang Nucleosynthesis. 

The present model uses complex scalars to realize the symmetry, and thus it 
involves dynamical phases for which we have not explored yet their possible 
role in the Universe evolution. Our analysis does show that they are not 
potentially relevant for the after-inflation evolution, provided the initial 
conditions fix them to zero, at least for the mechanism that generates 
neutrino masses and the production of quintessence quanta. However, other roles 
may be possible that would be interesting to look at.

Our analysis also explores the thermal corrections on the mass of $\Q$ coming 
from interactions with the thermal bath of neutrinos permeating the Universe.  
We found, by calculating the slow-roll condition under these corrections,  
that the $\Q$-field preserves its DE behaviour. At the present epoch, the 
thermal mass coincides with that commonly required in $\Q$-models.

As final comments. One of the issues that remain to be explored in detail to make
for a more realistic model is the phenomenology concerning to the $SO(1,1)$
flavor sector, above introduced as the connection
of our heavy neutrinos with the standard particle physics.
We have shown how heavy neutrinos and standard model particles should be assigned
into a set of doublets and singlets under $SO(1,1)$, thus, it would be interesting
to explore if such symmetry may account for masses and mixings of the light neutrinos as well.

\section*{Acknowledgments}
This work was done in part under financial support from Conacyt, Mexico, under 
grant No. 237004.

\appendix
\section{Diagonalization of the Lagrangian}

In this appendix, we present in detail the diagonalization analysis of our  
model Lagrangian whose results are used along the discussion in the main text.
First, we consider the scalar sector, whose Lagrangian 
(\ref{eq:scalar_sector}) in terms of the doublet complex field components 
becomes
$$
\Lag_{\Phi} = \partial^\mu\phi^*\partial_\mu\phi + 
\partial^\mu\varphi^*\partial_\mu\varphi - V(\phi, \varphi),
$$
with the potential
\begin{multline}\label{eq:potential_fields}
V(\phi, \varphi) = \alpha_0 \big(    |\phi|^2 + |\varphi|^2    \big) + \alpha_1 
\big(  \phi^* \varphi + \varphi^* \phi  \big)\\ + \alpha_3 \big(   \phi^2 - 
\varphi^2  \big) + c.c.
\end{multline}
Next, we  rewrite the Lagrangian in terms of the hermitian base
$$
\phi = \dfrac{1}{\sqrt 2} ( \phi_1 + i\phi_2), \qquad \varphi = \dfrac{1}{\sqrt 
2} (\varphi_1 + i\varphi_2),
$$
where $\phi_i,\ \varphi_i, \ i=1,2$ are real scalar fields.  This lets us put 
the potential in a matrix form which we will diagonalize in order to identify 
physical fields having separated dynamics. The potential 
(\ref{eq:potential_fields}) becomes
$$
V = \dfrac{1}{2}\Phi_R^\mathsmaller{T}A\Phi_R,
$$
with $\Phi_R$ being the vector formed from above real scalar fields components 
of $\phi$ and $\varphi$,  given by
$\Phi_R^\mathsmaller{T} = (\phi_1, \phi_2, \varphi_1, 
\varphi_2) $, 
and $A$ is the $4 \times 4$ mass coupling matrix
$$
A =
\begin{pmatrix}
  m_1^2     &  \lambda^2 & \mu_1^2 &   0   \\
  \lambda^2 &  m_2^2     &  0    & \mu_1^2 \\
\mu_1^2   & 0 & m_2^2  & -\lambda^2  \\
0  & \mu_1^2 & -\lambda^2 & m_1^2
\end{pmatrix},
$$
where we have defined
$$
m_1^2 = \mu_0^2 + \mu_3^2, \quad m_2^2 = \mu_0^2 - \mu_3^2,\quad \lambda^2 = 
2Re(i\alpha_3),
$$
and
$$
\mu_0^2 \equiv 2Re(\alpha_0),\quad\mu_1^2 \equiv 
2Re(\alpha_1), \quad \mu_3^2 = 2Re(\alpha_3)~.
$$
Notice that by definition  all the involved mass terms, $m_1^2$, 
$m_2^2$, $\lambda^2$, $\mu_0^2$, $\mu_3^2$ and $\mu_1^2$ are real and by 
construction, we have chosen them to be positive.

Since the $A$ matrix is real and symmetric, by mean of the proper orthogonal 
rotation of the field base, $\mathbb S$,  through which we redefine
$$
 \Phi_D = \mathbb S\Phi_R, \qquad A_D = \mathbb SA\mathbb S^{\mathsmaller T},
$$
we should get a diagonal mass sector. It is not difficult to check that such 
a matrix can be expressed as
$$
\mathbb S = \big( \mathbb I_{2\times 2} \otimes \mathbb B - i\sigma_2 \otimes 
\mathbb H  \big)\cos(\omega),
$$
where
$$
\mathbb B = \begin{pmatrix}  \cos(\rho) & 0 \\ 0 & \cos(\rho)   \end{pmatrix}, 
\qquad \mathbb H = \begin{pmatrix}  \tan(\omega) & \sin(\rho) \\ \sin(\rho) & 
-\tan(\omega)  \end{pmatrix}~.
$$
In the above, we have made use of the shorthand notation where
$$
\cos(\rho) = \dfrac{\mu_1^2}{\sqrt{\mu_1^4 + \lambda^4}}, \qquad \sin(\rho) = 
\dfrac{\lambda^2}{\sqrt{\mu_1^4 + \lambda^4}}~,
$$
$$
\cos(\omega) = \dfrac{\alpha^2}{\sqrt{2h^2(h^2 + \Delta^2)}}, \qquad 
\sin(\omega) = \dfrac{\alpha^2}{\sqrt{2h^2(h^2 - \Delta^2)}},
$$
and
$$
\alpha^4 = 4(\mu_1^4 + \lambda^4), \qquad \Delta^2 = m_1^2 - m_2^2, \qquad h^4 
= 
\Delta^4 + \alpha^4.
$$
After performing the $\mathbb S$ rotation,  the potential becomes 
$$
V = \dfrac{1}{2}\Phi_D^{\mathsmaller T} A_D \Phi_D, 
$$
with $\Phi_D^{\mathsmaller T} =(\Q_1 ,\xi_1 , \xi_2 , \Q_2)^{\mathsmaller T} $ and 
$$
A_D = diag
\begin{pmatrix}
  m^2, &  M^2, & M^2, & m^2 
\end{pmatrix},
$$
where the eigenvalues $m^2$ and $M^2$ are given by
\begin{equation}
m^2 = \mu_0^2 - \mu^2
\quad\text{and }\quad
M^2= \mu_0^2 + \mu^2~,  
\end{equation}
where $\mu^2 = \sqrt{\mu_3^4 + \mu_1^4 + \lambda^4}$. 
In terms of the $\alpha$ couplings, we get $\mu_0^2 = 2Re~\alpha_0$ and 
$\mu^2 = 2\sqrt{(Re~\alpha_1)^2 + |\alpha_3|^2}$~.

The requirement that $M^2, m^2 > 0$, which guarantees 
that the potential is bounded from below, is fulfilled if
$\mu_0^2 > \mu^2 >0$. If both parameters were of the same order, 
$\mu_0^2 \approx \mu^2 >0$, we would naturally get 
$M^2 \gg m^2\approx 0$. In such a scenario it 
becomes natural  to identify $\xi$ with the inflaton and $\Q$ with 
the DE field,  provided $M$ is as large as the inflation scale.

Notice that the  mass eigenstates in $\Phi_D$ can be 
rearranged in a  more natural ordering by the permutation matrix
$$
P =
\begin{pmatrix}
  1 &  0 & 0 &  0   \\
  0 &  0 & 0 &  1   \\
  0 &  1 & 0 &  0   \\
  0 &  0 & 1 &  0    
\end{pmatrix}~,
$$
such that $(\Q_1 , \Q_2, \xi_1 , \xi_2 )^T= \mathbb 
S^\prime\Phi_R$ with 
$\mathbb S^\prime = P\mathbb S$~.

In terms of the diagonal base and given that there are two degenerated 
scalar degrees of freedom for each mass, the potential finally can be expressed 
as
\begin{equation}\label{eq:potential_chi_xi}
V = m^2|\Q|^2 + M^2|\xi|^2,
\end{equation}
where we have introduced the new complex scalar fields
\begin{equation}\label{eq:chi_xi}
\Q = \dfrac{1}{\sqrt 2}(\Q_1 + i\Q_2), \qquad \text{and} \qquad \xi = 
\dfrac{1}{\sqrt 2}(\xi_1 + i\xi_2).
\end{equation}
Analogously, the scalar kinetic term can be easily put in terms 
of the new fields after the $\mathbb S^\prime$ rotation on $\Phi_R$, to get the 
also diagonal terms 
$ \partial^\mu\Q^*\partial_\mu\Q + \partial^\mu\xi^*\partial_\mu\xi$.

Finally, by introducing the doublet
\begin{equation}\label{eq:X_doublet}
  \rbvarphi =
  \begin{pmatrix}
    \Q \\ \xi    
  \end{pmatrix},
\end{equation}
the whole Lagrangian of the scalar sector becomes
\begin{equation}\label{eq:Lag_X}
\Lag_{\rbvarphi} = 
\partial^\mu\rbvarphi^\dagger\partial_\mu\rbvarphi - \rbvarphi^\dagger\mathbb M 
\rbvarphi,
\end{equation}
where $\mathbb M$ is the diagonal mass matrix
\begin{equation}\label{eq:mass_matrix_app}
\mathbb M =
\begin{pmatrix}
m^2 & 0 \\  0 & M^2
\end{pmatrix}.
\end{equation}
We should emphasize that this new doublet notation is not a faithful 
representation of $SO(1,1)$, since the $SO(4)$ rotation, $\mathbb S^\prime$,
and the $SO(1,1)$ transformations do not commute. Therefore, the diagonal 
Lagrangian (\ref{eq:mass_matrix_app}), which provides the decoupled field 
system which evolves explaining inflation and the 
late accelerated expansion of the Universe,  is not explicitly invariant under 
$SO(1,1)$, even though the original model does is so. 

Let us now move into analyzing the fermion sector of the theory, for which the 
corresponding kinetic terms, as given in Eq.~(\ref{eq:fermion_kinetic_terms}), 
are
\begin{equation}\label{eq:Weyl_kinetic}
 \Lag_{N_i} = \sum_{i=0}^{2} N_i^{\dagger a} i \sigma_{a\dot c}^\mu\partial_\mu 
N_i^{\dot c}~,
\end{equation}
and the interaction terms (\ref{eq:interaction_Phi}) which takes the form
\begin{multline}\label{eq:interaction_fields}
-\Lag_I = N_{0\dot a}\big\{ a_0(\phi^*N_1^{\dot a} + \varphi^*N_2^{\dot a}) +  
a_1(\phi^*N_2^{\dot a} + \varphi^*N_1^{\dot a})\\  +  a_2(\phi N_2^{\dot a} - 
\varphi N_1^{\dot a}) + a_3(\phi N_1^{\dot a} - \varphi N_2^{\dot a}) \big\} + 
h.c.
\end{multline}
Last, written in terms of the real field components in $\Phi_R$, leads to
\begin{equation}\label{eq:interaction_Phi_R}
-\Lag_I  = \dfrac{1}{\sqrt 2} N_{0\dot a}\Phi_R^\mathsmaller{T}\big\{ {\mathbb 
V}N_1^{\dot a} + {\ddGamma}{\mathbb V}N_2^{\dot a}  \big\} + h.c.,
\end{equation}
where $\mathbb V$ is the vector formed from the complex couplings $a_i$, given 
by
$$
\mathbb{V} =
\begin{pmatrix}
  a_3 + a_0 \\ i(a_3 - a_0)\\ a_1 - a_2\\ -i(a_1 + a_2)
  \end{pmatrix},
$$
and $\ddGamma$ is a $4 \times 4$ matrix given by
$\ddGamma = -\sigma_1\otimes\sigma_2$~.
After the $\mathbb S$ rotation in the scalar sector is set in, and noticing 
that $\ddGamma$ is actually an invariant matrix, since
$\ddGamma = \mathbb S \ddGamma \mathbb S^\mathsmaller{T}$, 
the interaction Lagrangian becomes
\begin{equation}\label{eq:interaction_Phi_D}
-\Lag_I  = \dfrac{1}{\sqrt 2} N_{0\dot a}\Phi_D^\mathsmaller{T}\left\{ {\mathbb 
V'}N_1^{\dot a} + {\ddGamma}{\mathbb V'}N_2^{\dot a}  \right\} + h.c.,
\end{equation}
where $\mathbb V' = \mathbb S \mathbb V$.

It is important to note that $\mathbb V'$ just corresponds to a redefinition of 
the Yukawa couplings, for which one can always assume a convenient 
parameterization, implicitly defined in terms of the initial 
$a_{i=0,\dots, 3}$ couplings. Hence, using this freedom we choose the following 
combinations to define the couplings in the rotated scalar base:
\begin{equation}\label{eq:new_couplings}
\mathbb{V'} = \dfrac{1}{\sqrt 2}
\begin{pmatrix}
  g_1 + g_2 \\ h_1 - h_2 \\ -i(h_1 + h_2) \\ i(g_1 - g_2)
  \end{pmatrix},
\end{equation}
where $g_{i=1,2}$ and  $h_{i=1,2}$ are complex numbers. Substituting the last 
expression and the redefinition of the scalar fields given in 
Eq.~(\ref{eq:chi_xi}) into  Eq.~(\ref{eq:interaction_Phi_D}), after some 
simple algebra, we finally rewrite the interaction terms as
\begin{multline}\label{eq:interaction_chi_xi}
-\Lag_I = N_{0\dot a}\{ g_1\Q F_1^{\dot a} + g_2\Q^* F_2^{\dot a} \\ + h_1\xi^* 
F_1^{\dot a} - h_2\xi F_2^{\dot a}  \} + h.c.,
\end{multline}
where the new Weyl fields $F_{i=1,2}^{\dot a}$ are the components of the doublet
\begin{equation}\label{eq:F_doublet}
  \mathbf F =
  \doublet{
    F_1^{\dot a} \\ F_2^{\dot a}
  },
\end{equation}
which in turn comes from the transformation
\begin{equation}\label{eq:Psi_to_F}
e^{-i\sigma_2 \pi/4}\Psi = \mathbf F,  
\end{equation}
i.e., the diagonalization of the scalar potential through $\mathbb S$, induces 
an $SO(2)$ rotation over the doublet Eq.~(\ref{eq:neutrino_doublet}), by an 
angle of $\pi/4$.  Note that we still can define the $U(1)$ 
global transformation used in (\ref{eq:U_1_transformation}) with the same 
charge for the new Weyl fields as $\mathbf F \longrightarrow e^{iq}\mathbf F$, 
and so this convenient transformation does not alter the argument used to 
remove the mass of $N_0$ in the main text. Nevertheless, as for the scalar 
sector, the transformations used to rewrite the interactions hide the $SO(1,1)$ 
symmetry of the theory, but on the other hand, allows to write down 
Eq.~(\ref{eq:interaction_chi_xi}) in a simple and compact way,  as
\begin{equation}\label{eq:interaction_F}
-\Lag_I = N_{0\dot a}\{ \rbvarphi^\dagger\mathbb G_1\mathbf F + 
\rbvarphi^\mathsmaller{T}\mathbb G_2\mathbf F \} + h.c.,
\end{equation}
where we have defined the coupling matrices as
\begin{equation}\label{eq:couplings_matrices_app}
\mathbb G_1 =
\begin{pmatrix}
0 & g_2 \\ h_1 & 0
\end{pmatrix}
\qquad
\mathbb G_2 =
\begin{pmatrix}
g_1 & 0 \\ 0 & -h_2
\end{pmatrix}.
\end{equation}
Finally, notice that  the transformation given in Eq.~(\ref{eq:Psi_to_F}) keeps 
the diagonal form of fermion kinetic terms, as expected, which can now
be expressed as
\begin{equation}\label{eq:fermion_kinetic_terms_2}
\Lag_{F} = N_0^{\dagger a} i \sigma^\mu_{a\dot c}\partial_\mu N_0^{\dot c} + 
\mathbf F^\dagger i \sigma^\mu\partial_\mu\mathbf F~.
\end{equation}

\section{Including phase fields on the $SO(1,1)$ model} 
\label{sec:appen_parametrization}

Here we explore some of the possible effects that 
considering dynamical phase fields for the cosmological scalars may have in the 
model outcomes discussed in the main text, as well as other interesting 
aspects that we believe might be of further interest for field dynamics. 
For this, we assume that after reheating, the  $\Q$ field remains dynamically  
trapped in a homogeneous and isotropic false vacuum configuration, which 
sources DE and breaks the $U(1)$ global symmetry in the neutrino sector,
whereas the inflaton field $\xi$ has already settled on its null value, 
and thus, quantum perturbation for our cosmological scalar 
fields can be conveniently introduced in a polar base as
\begin{equation}\label{eq:polar_base}
 \Q = \dfrac{(\langle \Q\rangle + \nchi)}{\sqrt 2}e^{i\vartheta/\langle 
\Q\rangle},\qquad 
 \xi = \dfrac{1}{\sqrt 2}|\xi| e^{i\theta/\langle \Q\rangle}, 
\end{equation}
where the degrees of freedom of the complex scalar field $\Q$ are now given by 
the real scalar field $\nchi$,  and the dynamical phase $\vartheta$. 
Similarly, for $\xi$, its degrees of freedom are given by its modulus and its 
own dynamical phase $\theta$.

Next, we proceed to rewrite the Lagrangian of our model in terms of the
above parameterization, for this we first notice that  the doublet  
(\ref{eq:X_doublet}) can be written as
\begin{equation}\label{eq:X_doublet_P_X_R}
\rbvarphi = \mathbb P \rbvarphi_{\mathsmaller R},
\end{equation}
where we have defined the radial field part as 
\begin{equation}\label{eq:X_R}
  \rbvarphi_{\mathsmaller{R}} =
\dfrac{1}{\sqrt 2}
\doublet{
\langle \Q\rangle + \nchi \\ |\xi|
},
\end{equation} 
and the field phase matrix given by
\begin{equation}
  \mathbb P =
  \begin{pmatrix}
    e^{i\vartheta/\langle \Q\rangle} & 0 \\
    0 & e^{i\theta/\langle \Q\rangle}
  \end{pmatrix}.
\end{equation}
By substituting Eq.~(\ref{eq:X_doublet_P_X_R}) into the scalar sector of the 
theory, it is straightforward to see that the Lagrangian (\ref{eq:Lag_X}) 
simply becomes
\begin{equation}\label{eq:Lag_rbvarphi_R_T}
\Lag_{\rbvarphi} =  \partial^\mu\rbvarphi^{\mathsmaller 
T}_{\mathsmaller R} \partial_\mu\rbvarphi_{\mathsmaller R} +  
\rbvarphi^{\mathsmaller T}_{\mathsmaller R}\mathbb M \rbvarphi_{\mathsmaller R} 
+   \mathcal T(\rbvarphi_{\mathsmaller R},\mathbb P)~,
\end{equation}
where $\mathcal T(\rbvarphi_{\mathsmaller R},\mathbb P) =   
\rbvarphi^{\mathsmaller 
T}_{\mathsmaller R}\left(\partial^\mu\mathbb 
P^\dagger\right)\left(\partial_\mu\mathbb P\right)\rbvarphi_{\mathsmaller R}$, 
is a dimension six and highly suppressed operator. Thus we do not expect it to 
be relevant for the later dynamics of DE.

Explicitly, in terms of inflaton and DE fields, the above Lagrangian reads
\begin{multline}\label{eq:scalar_fields_Lag_final}
\Lag_{\rbvarphi} =   \dfrac{1}{2} 
\partial^\mu|\xi|\partial_\mu|\xi| + \dfrac{1}{2}\partial^\mu 
\nchi\partial_\mu \nchi\\ 
  + \dfrac{1}{2} m^2 (\langle \Q\rangle + \nchi)^2 + \dfrac{1}{2} M^2 |\xi|^2 + 
\mathcal 
T_{(\xi,\nchi,\vartheta,\theta)},
\end{multline}
where the last term on the RHS is given by
\begin{equation}\label{eq:scalar_currents_X_xi}
\mathcal T_{(\xi,\nchi,\vartheta,\theta)} = 
\dfrac{|\xi|^2}{2\langle \Q\rangle^2}\partial^\mu\theta\partial_\mu\theta  +
 \dfrac{1}{2}\left( 1 + \dfrac{\nchi}{\langle \Q\rangle} 
\right)^2\partial^\mu\vartheta\partial_\mu\vartheta.
\end{equation}

As for the interaction with fermions given by Eq.~(\ref{eq:interaction_F}), 
this is now written as
\begin{eqnarray}\label{eq:Lag_X_R_A}
-\Lag_I &=& N_{0\dot a}\rbvarphi_{\mathsmaller{R}}^{\mathsmaller{T}}\left\{ 
\mathbb P^\dagger \mathbb G_1 + \mathbb P^{\mathsmaller T}\mathbb G_2 \right\} 
\mathbf F + h.c., \\
\label{eq:Lag_X_R}
&=& N_{0\dot a}\rbvarphi_{\mathsmaller{R}}^{\mathsmaller{T}}\mathbb G 
\mathbf F' + h.c.,
\end{eqnarray}
where the new coupling matrix is given by
\begin{equation}\label{eq:G_matrix}
\mathbb G =
\begin{pmatrix}
g_1 & g_2 \\ h_1 e^{-i(\theta + \vartheta)/\langle \Q\rangle} & -h_2e^{i(\theta 
+ 
\vartheta)/\langle \Q\rangle}
\end{pmatrix}~,
\end{equation}
and where we have performed a local phase transformation over the fermions in 
the doublet to introduce 
\begin{equation}\label{eq:F_prim_doublet}
  {\mathbf F'} =
  \doublet{
    F_1^{'\dot a} \\ F_2^{'\dot a}
  }~,
\end{equation}
with
$F_1^{'\dot a} = e^{i\vartheta/\langle \Q\rangle}F_1^{\dot a}$ and 
$F_2^{'\dot a} = e^{-i\vartheta/\langle \Q\rangle}F_2^{\dot a}$.
This redefinition of the fermion fields removes the dynamical phases on the
$\nchi$-sector, as can be seen from (\ref{eq:G_matrix}). Nonetheless,   they 
will 
reappear as currents coming from the transformation of the kinetic terms 
(\ref{eq:fermion_kinetic_terms_2}), which now read as
\begin{equation*}
\Lag_{F} = N_0^{\dagger a} i \sigma^\mu_{a\dot c}\partial_\mu N_0^{\dot c} + 
\mathbf F'^\dagger i \sigma^\mu\partial_\mu\mathbf F' + 
\dfrac{\partial_\mu\vartheta}{\langle \Q\rangle}\mathbf F'^\dagger 
\sigma^\mu\sigma_3\mathbf 
F',
\end{equation*}
where in the last term the effect of  $\sigma_3$  is to switch the sign of the 
lower entry of the doublet. Notice that once again the phase field enters in a 
suppressed way. Apart from these new terms where the phase fields are 
explicit, the part of the Lagrangian that matters for the model remains the 
same.

\subsection{Revisiting massive neutrino base}

Let us now execute a new transformation with the aim to remove the constant 
phases of the couplings $g_1$ and $g_2$ appearing in (\ref{eq:G_matrix}), 
by means of a $SU(2)$ rotation on the doublet fermion sector
\begin{equation}\label{eq:doublet_eta}
  {\large{\boldsymbol \eta}} =
  \mathbb R \mathbf F' = 
  \doublet{
    \eta_1^{\dot a} \\ \eta_2^{\dot a}
  },
\end{equation}
with
\begin{equation}\label{eq:R_matrix}
\mathbb R = \dfrac{1}{a_c}
\begin{pmatrix}
g_1 & g_2 \\ -g_2* & g_1*
\end{pmatrix},
\end{equation}
where 
$a_c = \sqrt{|g_1|^2 + |g_2|^2}$. After this rotation, the interaction term 
(\ref{eq:Lag_X_R}) becomes
\begin{equation}\label{eq:Lag_n}
-\Lag_I = N_{0\dot a}\rbvarphi_{\mathsmaller{R}}^{\mathsmaller{T}}\mathbb G' 
{\large{\boldsymbol \eta}} + h.c.,
\end{equation}
where now, the coupling matrix is
\begin{equation}\label{eq:G_prime_matrix}
\mathbb G' = \mathbb G \mathbb R^\dagger =
\begin{pmatrix}
a_c & 0 \\ C_1(\theta,\vartheta)  &  C_2(\theta,\vartheta)
\end{pmatrix}~.
\end{equation}
In above we have used for a shorthand notation 
\begin{equation*}
\begin{split}
  C_1(\theta,\vartheta) &= \ \ (g_{11} e^{-i(\theta + \vartheta)/
  \langle \Q\rangle} - g_{22} 
e^{i(\theta + \vartheta)/\langle \Q\rangle}) / a_c,\\ 
  C_2(\theta,\vartheta) &= -(g_{12} e^{i(\theta + \vartheta)/\langle \Q\rangle} 
+ g_{21} 
e^{-i(\theta + \vartheta)/\langle \Q\rangle}) / a_c,
\end{split}
\end{equation*}
where 
$g_{11} = g_1^* h_1$, $g_{22} = g_2^* h_2$, $g_{12} = g_1 h_2$,
and $g_{21} = g_2 h_1$.
On the other hand, upon the same rotation, the fermion kinetic terms are now 
written as
\begin{equation}\label{eq:kinetic_eta_prime}
\Lag_{F} = N_0^{\dagger a} i\sigma_{a\dot c}^\mu\partial_\mu N_0^{\dot c} + 
{\large{\boldsymbol \eta}}^\dagger i\sigma^\mu\partial_\mu{\large{\boldsymbol 
\eta}} + \dfrac{\partial_\mu\vartheta}{\langle\Q\rangle}{\large{\boldsymbol \eta}}^\dagger 
\sigma^\mu \mathbb Y {\large{\boldsymbol \eta}},
\end{equation}
where $\mathbb Y$ is a couplings matrix, that comes from the transformation of 
$\sigma_3$ under (\ref{eq:R_matrix}), given by
$$
\mathbb Y =
\begin{pmatrix} y_1 & -y_2 \\ -y_2^* & -y_1 \end{pmatrix},
$$
where $y_1 = \left(|g_1|^2 - |g_2|^2\right)/a_c^2$, and $y_2 = 2g_1g_2/a_c^2$, 
i.e., $y_1 \in \Reals$ and $y_2 \in \Complex$.
(Notice that  $y_1^2 + |y_2|^2 = 1$.)

Let us now concentrate our analysis towards the interaction among neutrinos 
and the DE field, which after above mathematical manipulations has gotten the 
simple expression
\begin{equation}\label{eq:Lag_nu_Q}
-\Lag_{\nu \nchi} = \dfrac{a_c}{\sqrt 2}(\langle \Q\rangle + 
\nchi)\left\{ N_{0\dot 
a}\eta_1^{\dot a} + h.c.\right\}.
\end{equation}
The part between braces can be expressed also as
\begin{align}\label{eq:Lag_nu_Q_part}
  N_{0\dot a}\eta_1^{\dot a} + h.c. & =  N_{0\dot a}\eta_1^{\dot a} + 
\eta_1^{\dagger a}N_{0a}^\dagger \\ \notag
  & = \dfrac{1}{2}\{ N_{0\dot a}\eta_1^{\dot a} + N_{0\dot a}\eta_1^{\dot a} + 
\eta_1^{\dagger a}N_{0a}^\dagger + \eta_1^{\dagger a}N_{0a}^\dagger \}\\ \notag
  & = \dfrac{1}{2}\{ N_{0\dot a}\eta_1^{\dot a} + \eta_{1\dot a}N_0^{\dot a} + 
\eta_1^{\dagger a}N_{0a}^\dagger + N_0^{\dagger a}\eta_{1a}^\dagger\}, \notag
\end{align}
wherein both, the second and the fourth terms in the last line, we have used 
the anti-commutation properties plus an extra minus sign coming from the change 
from $^{\dot a}\,_{\dot a}$ to  $_{\dot a}\,^{\dot a}$ (and similarly for the 
undotted indices). Now, we define two four-component Dirac neutrinos as
\begin{equation}\label{eq:u_1_and_u_2}
  u_1 =
  \doublet{
    N_{0a}^\dagger \\ \eta_1^{\dot a}
  },
  \qquad
  u_2 =
  \doublet{
    \eta_{1a}^\dagger \\ N_0^{\dot a}
  },
\end{equation}
in terms of which  the last line in Eq.~(\ref{eq:Lag_nu_Q_part}) can be written 
as
\begin{equation}\label{eq:Lag_nu_Q_part_2}
N_{0\dot a}\eta_1^{\dot a} + h.c. = \dfrac{1}{2}\{ \bar u_1 u_1 + \bar u_2 u_2  
\}.
\end{equation}

As it can be seen from (\ref{eq:u_1_and_u_2}), the neutrinos $u_1$ and $u_2$ are 
conjugates of charge of each other, this let us put them in terms of two 
Majorana neutrinos $\nu_1$ and $\nu_2$, through of another rotation, which is 
given by
\begin{equation}\label{eq:u_i_to_nu_i}
\begin{pmatrix} \nu_1 \\ \nu_2  \end{pmatrix} =
\dfrac{1}{\sqrt 2}
\begin{pmatrix} 1 & 1 \\ -i & i  \end{pmatrix}
\begin{pmatrix} u_1 \\ u_2      \end{pmatrix}.
\end{equation}
Therefore, Eq.~(\ref{eq:Lag_nu_Q_part_2}) directly becomes
\begin{equation}\label{eq:Lag_nu_Q_part_3}
N_{0\dot a}\eta_1^{\dot a} + h.c. = \dfrac{1}{2}\{ \bar\nu_1 \nu_1 + \bar \nu_2 
\nu_2  \},
\end{equation}
which explicitly provide the neutrino mass eigenstates, with a mass given 
by
\begin{equation}\label{eq:neutrino_mass_def_app}
m_k = \dfrac{a_c\langle \Q\rangle}{\sqrt 2}~.
\end{equation}
Notice that this same rearrangement of the neutrinos 
provide the interaction Lagrangian with $\nchi$ fields,
\begin{equation}
-\Lag_{I \nchi} = \dfrac{a_c}{2\sqrt 2}\nchi\left( 
\bar\nu_1\nu_1 + \bar\nu_2\nu_2\right),
\end{equation}
that we use  on our discussions along the paper.
We stress that these results are independent of the phase fields and link the 
origin of the heavy right handed neutrino masses to DE, as already argued 
in the main text.

As a final note on this regard, notice that the Majorana neutrinos, in 
four-component notation, can be  expressed as
\begin{equation}\label{eq:nu_Majorana}
  \nu_i =
  \doublet{
   \mathtt K_{ia}^\dagger \\ \mathtt K_i^{\dot a}
  },\quad i=1,2. 
\end{equation}
In the last equation, we have introduced the new right-handed Weyl field in 
two-component notation: $\mathtt K_{i=1,2}^{\dot a}$.
Note that the transformation (\ref{eq:u_i_to_nu_i}) together with 
(\ref{eq:u_1_and_u_2}) are equivalent to the transformations
\begin{equation}\label{eq:OMR_1}
  \doublet{
    \mathtt K_1^{\dot a} \\ \mathtt K_2^{\dot a}
  } =
  \dfrac{1}{\sqrt 2}
  \begin{pmatrix} 1 & 1 \\ i & -i \end{pmatrix}
  \doublet{ N_0^{\dot a} \\ \eta_1^{\dot a}
  },
\end{equation}
and
\begin{equation}\label{eq:OMR_2}
  \doublet{
    \mathtt K_{1a}^\dagger \\ \mathtt K_{2a}^\dagger
  } =
  \dfrac{1}{\sqrt 2}
  \begin{pmatrix} 1 & 1 \\ i & -i \end{pmatrix}
  \doublet{
    \eta_{1a}^\dagger \\ N_{0a}^\dagger
  }.
\end{equation}
It is important to remark that these transformations do not respect  
the  $U(1)$ invariance of the fermionic sector since it mixes fields with 
different global charges.

Summarizing, we can either, substitute (\ref{eq:nu_Majorana}) into  
(\ref{eq:Lag_nu_Q_part_3}) or directly operate over (\ref{eq:Lag_nu_Q_part}) 
through of (\ref{eq:OMR_1}) and (\ref{eq:OMR_2}) to get
\begin{equation}\label{eq:Lag_nu_Q_part_4}
  N_{0\dot a}\eta_1^{\dot a} + h.c. = \dfrac{1}{2}\left\{\mathtt K_{1\dot 
a}\mathtt K_1^{\dot a} + \mathtt K_{2\dot a}\mathtt K_2^{\dot a}\right\} + h.c.
\end{equation}
By substituting equation (\ref{eq:Lag_nu_Q_part_4}) into equation 
(\ref{eq:Lag_nu_Q}), one gets  the mass terms 
\begin{equation}\label{eq:Lag_mk}
-\Lag_m = \dfrac{1}{2}m_k\left(\mathtt K_{1\dot a}\mathtt K_1^{\dot a} + 
\mathtt 
K_{2\dot a}\mathtt K_2^{\dot a}\right) + h.c.,
\end{equation}
with the mass given as before and the interaction term
\begin{equation}\label{eq:Lag_IQ}
-\Lag_{I \nchi} = \dfrac{a_c}{2\sqrt 2}\nchi\left(\mathtt K_{1\dot a}\mathtt 
K_1^{\dot a} + \mathtt K_{2\dot a}\mathtt K_2^{\dot a}\right) + h.c.
\end{equation}

In the same footing, and for future use, we also write  
the inflaton to neutrino interactions, as derived from Eq.~(\ref{eq:Lag_n}), 
for which we also rename $\mathtt K_3^{\dot a}  \equiv \eta_2^{\dot a}$, to 
write
\begin{multline}\label{eq:Lag_g_K_i}
    -\Lag_g = \dfrac{1}{4}C_1(\theta,\vartheta)|\xi|\left(\mathtt K_{1\dot 
a}\mathtt K_1^{\dot a} + \mathtt K_{2\dot a}\mathtt K_2^{\dot a}\right)\\
    + \dfrac{1}{2\sqrt 2}C_2(\theta,\vartheta)|\xi|\left(\mathtt K_{1\dot a} - 
i\mathtt K_{2\dot a}\right)\mathtt K_3^{\dot a}  + h.c.
\end{multline}
Similarly, by expanding Eq.~(\ref{eq:kinetic_eta_prime}) and by 
transformation  (\ref{eq:OMR_1}), whereas
the kinetic terms for $\mathtt K_{i=1,2,3}^{\dot a}$ remain as usual,
\begin{equation}\label{eq:Lag_K}
\Lag_K = \sum_{i=1}^3 \mathtt K_i^{\dagger a} i\sigma_{a\dot c}^\mu\partial_\mu 
\mathtt K_i^{\dot c},
\end{equation}
the current-couplings among the phase scalar $\partial_\mu\vartheta$ 
and the neutrinos go as
\begin{equation}
\Lag_c = \Lag_{c_1} + \Lag_{c_2},
\end{equation}
where
\begin{align}\label{eq:Lag_c_1}
  &\Lag_{c_1} = y_1\,\dfrac{\partial_\mu\vartheta}{\langle \Q\rangle}  \left\{  
\dfrac{1}{2}\left( \mathtt K_1^{\dagger a}\sigma_{a\dot c}^\mu \mathtt 
K_1^{\dot 
c}  +   \mathtt K_2^{\dagger a}\sigma_{a\dot c}^\mu \mathtt K_2^{\dot 
c}\right)\right.\notag\\
  &+ \left.\dfrac{i}{2}\left( \mathtt K_1^{\dagger a}\sigma_{a\dot c}^\mu 
\mathtt K_2^{\dot c} - \mathtt K_2^{\dagger a}\sigma_{a\dot c}^\mu \mathtt 
K_1^{\dot c}\right)  - \mathtt K_3^{\dagger a}\sigma_{a\dot c}^\mu \mathtt 
K_3^{\dot c}\right\},
\end{align}
and
\begin{equation}\label{eq:Lag_c_2}
\Lag_{c_2} = -\dfrac{\partial_\mu\vartheta}{\langle \Q\rangle}\left\{ 
\dfrac{y_2}{\sqrt 
2}(\mathtt K_1^{\dagger a} - i\mathtt K_2^{\dagger a})\sigma_{a\dot c}^\mu 
\mathtt K_3^{\dot c} + h.c.    \right\}.
\end{equation}

\subsection{Energy density and equations of motion for the DE sector}
We close this appendix by presenting the results of the calculation of 
the equation of state parameter for DE in the present model. 
For this purpose, we made explicit use of the model Lagrangian, as defined in 
Eq.~(\ref{eq:scalar_fields_Lag_final}),  where the DE part is written as
\begin{equation}\label{eq:Lag_X_vartheta_sep}
\Lag_{\nchi,\vartheta} = \dfrac{1}{2}\partial^\mu \nchi\partial_\mu \nchi + 
\dfrac{1}{2}\left( 1 + \dfrac{\nchi}{\langle \Q\rangle} 
\right)^2\partial^\mu\vartheta\partial_\mu\vartheta + V(\nchi), 
\end{equation}
where  the potential is defined as
\begin{equation}\label{eq:potential_X}
V(\nchi) =  \dfrac{1}{2} m^2 (\langle \Q\rangle + \nchi)^2.
\end{equation}
From equation (\ref{eq:Lag_X_vartheta_sep}) and by calculation of the
energy-momentum tensor in an FLRW Universe, we obtain both, the energy density and the 
pressure in terms of $\nchi$ and the phase $\vartheta$. These are given by
\begin{multline}\label{eq:energy_density_X_vartheta}
\rho_{\mathsmaller{DE}} = \dfrac{1}{2}\dot\nchi^2 + \dfrac{1}{2}\left( 1 + 
\dfrac{\nchi}{\langle \Q\rangle} \right)^2\dot\vartheta^2 + 
\dfrac{1}{2a^2}(\nabla\nchi)^2 \\ 
+ V(\nchi) + \dfrac{1}{2a^2}\left( 1 + 
\dfrac{\nchi}{\langle \Q\rangle}\right)^2(\nabla\vartheta)^2~,
\end{multline}
and
\begin{multline}\label{eq:pressure_X_vartheta}
P_{\mathsmaller{DE}} = \dfrac{1}{2}\dot\nchi^2 + \dfrac{1}{2}\left( 1 + 
\dfrac{\nchi}{\langle \Q\rangle} \right)^2\dot\vartheta^2 - 
\dfrac{1}{6a^2}(\nabla\nchi)^2 \\ 
- V(\nchi) - \dfrac{1}{6a^2}\left( 1 + 
\dfrac{\nchi}{\langle \Q\rangle}\right)^2(\nabla\vartheta)^2~.
\end{multline}
In the homogeneous case, the previous equations are reduced to
\begin{equation}\label{eq:energy_density_X_vartheta_homogen}
\rho_{\mathsmaller{DE}} = \dfrac{1}{2}\dot\nchi^2 + \dfrac{1}{2}\left( 1 + 
\dfrac{\nchi}{\langle \Q\rangle} \right)^2\dot\vartheta^2  + V(\nchi),
\end{equation}
and
\begin{equation}\label{eq:pressure_X_vartheta_homogen}
  P_{\mathsmaller{DE}} = \dfrac{1}{2}\dot\nchi^2 + \dfrac{1}{2}\left( 1 + 
\dfrac{\nchi}{\langle \Q\rangle} \right)^2\dot\vartheta^2  - V(\nchi).
\end{equation}
In order to realize the accelerated expansion, the DE field has to accomplish an 
equation of state such that
$$
\omega \equiv \dfrac{P_{\mathsmaller{DE}}}{\rho_{\mathsmaller{DE}}} \approx -1,
$$
which means, according to (\ref{eq:energy_density_X_vartheta_homogen}) and 
(\ref{eq:pressure_X_vartheta_homogen}), that the first slow-roll condition is of 
the form
\begin{equation}\label{eq:first_slow_roll_condition}
\dfrac{1}{2}\dot\nchi^2 + \dfrac{1}{2}\left( 1 + 
\dfrac{\nchi}{\langle \Q\rangle} 
\right)^2\dot\vartheta^2  \ll \dfrac{1}{2}m^2(\langle \Q\rangle + \nchi)^2.
\end{equation}

The dynamics for the homogeneous background involving both, $\nchi$ and 
$\vartheta$, is given by substitution of the equation 
(\ref{eq:energy_density_X_vartheta_homogen})  into the first Friedman equation, 
after application of the first slow-roll condition, together with those coming 
from application of the Euler-Lagrange equations to 
(\ref{eq:Lag_X_vartheta_sep}). For completeness, we also included DM, baryons 
($b$), photons ($\gamma$) and active neutrinos ($n$). Taking into account the 
first slow-roll condition, the whole system is: 
\begin{equation}\label{eq:dynamic_system_background}
  \begin{split}
  &\mathsf H^2 = \dfrac{1}{3M_{pl}^2}V(\nchi),\\ 
  &\ddot\nchi + 3\mathsf H\dot\nchi + V(\nchi)_{,\nchi} = 0,\\
  &\ddot\vartheta +  3\mathsf H\dot\vartheta = 0,\\
  &\mathsf{\dot H} = \dfrac{-1}{2M_{pl}^2}\left( \rho_{\mathsmaller{DM}} + \rho_b + 
\dfrac{4}{3}\rho_\gamma + \dfrac{4}{3}\rho_\nu \right),\\
  &\dot\rho_{\mathsmaller{DM},b} + 3\mathsf H\rho_{\mathsmaller{DM},b} = 0,\\
  &\dot\rho_{\gamma,n} + 4\mathsf H\rho_{\gamma,n} = 0.
\end{split}
\end{equation}
\newline


\end{document}